%% file: main.tex
\documentclass[10pt, conference, compsocconf]{IEEEtran}

\usepackage{mathptmx} 
\usepackage[table]{xcolor} 
\usepackage{arydshln}
\usepackage{fancyhdr}
\usepackage[normalem]{ulem}
\usepackage[hyphens]{url}
\usepackage[sort,nocompress]{cite}
\usepackage[final]{microtype}

\usepackage{CJKutf8}
\usepackage{pagecolor}
\usepackage{microtype}
\usepackage{graphicx}
\usepackage[caption=false]{subfig}
\usepackage{float}

\usepackage{booktabs} 
\usepackage{url}
\usepackage{multirow}
\usepackage{listings}
\usepackage{todo}

\usepackage{xspace}
\usepackage{amsmath,amssymb,amsfonts}
\usepackage{bm} 
\usepackage{lipsum}
\usepackage{tabularx}
\usepackage{adjustbox}
\usepackage{algorithm}
\usepackage{siunitx}
\usepackage{algorithmic}
\usepackage[bookmarks=true,pdfpagelabels=false,breaklinks=true,colorlinks,linkcolor=black,citecolor=blue,urlcolor=black]{hyperref}
\usepackage[acronym]{glossaries}
\usepackage[capitalise,nameinlink]{cleveref}
\Crefformat{figure}{#2Fig.~#1#3}
\Crefformat{table}{#2Tab.~#1#3}

\newcommand\LONG{}
\newcommand\tableninja{}

\ifdefined\LONG\else\fi

\usepackage[splitrule]{footmisc} 

\usepackage{mathtools, nccmath}
\DeclarePairedDelimiter{\nint}\lfloor\rceil
\makeatletter
\DeclareRobustCommand\bigop[2][1]{%
  \mathop{\vphantom{\sum}\mathpalette\bigop@{{#1}{#2}}}\slimits@
}
\newcommand{\bigop@}[2]{\bigop@@#1#2}
\newcommand{\bigop@@}[3]{%
  \vcenter{%
    \sbox\z@{$#1\sum$}%
    \hbox{\resizebox{\ifx#1\displaystyle#2\fi\dimexpr\ht\z@+\dp\z@}{!}{$\m@th#3$}}%
  }%
}
\makeatother

\newcommand{\RealNum}{\mathbb{R}}

\newcommand{\tensorDim}[5]{$\langle#1,#2,#3,#4,#5\rangle$}

\include{./symbols}

\definecolor{codeblue}{rgb}{0.2,0.2,1}
\definecolor{codegreen}{rgb}{0,0.5,0}
\definecolor{codegray}{rgb}{0.5,0.5,0.5}
\definecolor{codepurple}{rgb}{0.58,0,0.82}
\definecolor{backcolour}{rgb}{0.95,0.95,0.92}
\lstdefinestyle{mystyle}{
    commentstyle=\color{codegreen},
    numberstyle=\tiny\color{codegray},
    stringstyle=\color{codepurple},
    basicstyle=\scriptsize\ttfamily,
    breakatwhitespace=false,         
    breaklines=true,                 
    captionpos=b,                    
    keepspaces=true,                 
    numbers=left,                    
    numbersep=5pt,                  
    showspaces=false,                
    showstringspaces=false,
    showtabs=false,
    frame=single,
    tabsize=2,
    columns=flexible,
    keywordstyle=\color{codeblue}\bfseries,
    keywordstyle = [2]{\color{codepurple}},
    keywordstyle = [3]{\color{orange}},
    otherkeywords = {True, transfer, mmad, input_xform, weight_xform, out_xform, dconv, write},
    morekeywords = [2]{get_next_tile, transfer, mmad, input_xform, weight_xform, out_xform, dconv, write},
    morekeywords = [3]{broadcast, double_buffering},
}
\newcommand{\revised}[1]{#1}
\newcommand{\clarity}[1]{#1}

\newcommand{\shepherd}[1]{#1}

\begin{document}

\title{Going Further With Winograd Convolutions:\\ Tap-Wise Quantization for Efficient Inference on 4x4 Tiles}

\makeatletter
\newcommand{\linebreakand}{%
  \end{@IEEEauthorhalign}
  \hfill\mbox{}\par
  \mbox{}\hfill\begin{@IEEEauthorhalign}
}
\def\ps@IEEEtitlepagestyle{
  \def\@oddfoot{\mycopyrightnotice}
  \def\@evenfoot{}
}
\def\mycopyrightnotice{
  {\footnotesize
  \begin{minipage}{\textwidth}
  \centering
  Copyright~\copyright~2022 IEEE. Personal use of this material is permitted. Permission from IEEE must be obtained for all other uses, in any current or future media, including reprinting/republishing this material for advertising or promotional purposes, creating new collective works, for resale or redistribution to servers or lists, or reuse of any copyrighted component of this work in other works.
  \end{minipage}
  }
}
\makeatother

 \author{\IEEEauthorblockN{
     Renzo Andri\IEEEauthorrefmark{1}, 
     Beatrice Bussolino\IEEEauthorrefmark{1}\IEEEauthorrefmark{2}\textsuperscript{1}, 
     Antonio Cipolletta\IEEEauthorrefmark{1}, 
     Lukas Cavigelli\IEEEauthorrefmark{1},
     Zhe Wang\IEEEauthorrefmark{1}}
     \IEEEauthorblockA{\\\IEEEauthorrefmark{1}Computing Systems Lab, Huawei Zurich Research Center \\
     \IEEEauthorrefmark{2}DET, Politecnico di Torino, Italy}
     }

\maketitle


\begin{abstract}
Most of today's computer vision pipelines are built around deep neural networks, where convolution operations \clarity{require} most of the generally high compute effort.  
The Winograd convolution algorithm \clarity{computes convolutions with fewer \glspl{MAC} compared to the standard algorithm}, reducing the operation count by a factor of 2.25× for 3×3 convolutions when using the version with 2×2-sized tiles $F_2$. 
Even though the gain is significant, \clarity{the Winograd algorithm} with larger tile sizes, i.e., $F_4$, offers even more potential in improving throughput and energy efficiency, as it reduces the required \glspl{MAC} by 4×. 
Unfortunately, \clarity{the Winograd algorithm} with larger tile sizes introduces numerical issues that prevent its use on integer \glspl{DSA} and higher computational overhead to transform input and output data between spatial and Winograd domains.

To unlock the full potential of Winograd $F_4$, we propose a novel tap-wise quantization method that overcomes the numerical issues of using larger tiles, \clarity{enabling integer-only inference}.
\clarity{Moreover,} we present custom hardware units that process the Winograd transformations in a power- and area-efficient way, and we show how to integrate such custom modules in an \clarity{industrial-grade, programmable DSA}.
An extensive experimental evaluation on a large set of state-of-the-art computer vision benchmarks reveals that the tap-wise quantization algorithm \clarity{makes the quantized Winograd $F_4$ network almost as accurate as the FP32 baseline}. The Winograd-enhanced DSA achieves up to \revised{1.85×} gain in energy efficiency and up to \revised{1.83×} end-to-end speed-up for state-of-the-art segmentation and detection networks.
\end{abstract}

\begin{IEEEkeywords}
\shepherd{Machine Learning Acceleration, Winograd Convolution, ML System Design}

\end{IEEEkeywords}
\setcounter{footnote}{1}
\footnotetext{This work was done while Beatrice Bussolino was an intern at
Huawei Technologies - Zurich Research Center.}

\input{1_introduction}

\input{2_algorithm}
\input{3_hardwareAccel}
\input{4_ExperimantalEvaluation}
\input{5_relatedworks}
\input{6_conclusion}

\bibliography{andri_mendeley,references}
\bibliographystyle{IEEEtranS}

\end{document}

%% file: symbols.tex
\ifdefined\RELEASEON
    \newcommand\cipo[1]{}
    \newcommand\busso[1]{}
    \newcommand\cavi[1]{}
    
\else

\fi

\newcommand\smallEquationFont{\footnotesize}

\newcommand{\unit}[1]{\textit{#1}}

\newcommand{\x}{×}

\newlength\myindent
\newacronym{AI}{AI}{Artificial Intelligence}
\newacronym{GEMM}{MatMul}{matrix-matrix multiplication}
\newacronym{DFT}{DFT}{discrete Fourier transform}
\newacronym{DSA}{DSA}{domain-specific accelerator}
\newacronym{FM}{FM}{feature map}
\newacronym{iFM}{iFM}{input feature map}
\newacronym{oFM}{oFM}{output feature map}
\newacronym{MAC}{MAC}{multiply–accumulate operation}
\newacronym{CNN}{CNN}{convolutional neural network}
\newacronym{PE}{PE}{processing element}
\newacronym{WA-static}{WA-static}{Winograd-aware-static}

%% file: 1_introduction.tex
\section{Introduction}
\label{sec:introduction}

\Glspl{CNN} have had a significant breakthrough in almost all \clarity{Artificial Intelligence (AI)} tasks in recent years, thanks to large, comprehensible, and publicly available data sets, easy-to-use frameworks, and the newly available vast compute resources. 
Nevertheless, state-of-the-art neural networks come with high computational costs ($>$GFLOPs/inference) and memory requirements ($>$10 MB), which has led to an explosion of \glspl{DSA} in datacenters~\cite{TPU2016, liao2019davinci} and edge devices~\cite{kang2021benchmarking}.

Typically, floating-point numbers have been used to flexibly represent \gls{CNN} computations, but floating-point \clarity{datapath is} very area- and power-hungry due to the large intermediate values, re-normalization, and exception handling, which require adequate hardware support. 
Thanks to the intrinsic error tolerance of neural networks~\cite{zhu2020towards, guo2018survey}, 8-bit integer operations can be used for most inference applications with no to minimal accuracy degradation.
As integer operations---like additions and multiplications---are one order of magnitude more energy-efficient than their floating-point counterparts~\cite{horowitz2014energy}, integer-based accelerators can achieve much higher peak throughput and energy efficiency.

In order to reduce operation count and memory footprint even further, recent works have proposed to adopt smaller convolutional kernel sizes~\cite{Krizhevsky2012a}, exploit sparsity~\cite{Han2015}, use group convolutions~\cite{Krizhevsky2012a, xie2017aggregated}, channel shuffling~\cite{Zhang2017}, and depthwise separable convolutions~\cite{howard2017mobilenets}. 
Nevertheless, compute-heavy, dense $3{\times}3$ convolutional layers are still widely used in many state-of-the-art computer vision models~\cite{ssd300_vgg16, redmon2018yolov3, unet}. 
\clarity{Thus, the adoption} of the Winograd algorithm~\cite{winograd1980arithmetic} represents an interesting \clarity{optimization} opportunity as it converts the $3{\times}3$ convolution operation into a much less expensive elementwise multiplication.


The Winograd convolution algorithm extends the Toom-Cook algorithm to support convolutions by applying the (polynomial) Chinese remainder theorem, minimizing the number of required multiplications \cite{winograd1980arithmetic,blahut2010fast}.
Specifically, the 2D convolution with a feature map $x$ of size $m{\times}m$ and convolutions kernel $f$ of size $3{\times}3$ is calculated with the Winograd (convolution) algorithm $F_m = F(m{\times}m, 3{\times}3)$ as follows: 
\begin{align}\label{eq:winograd}
Y=A^T \left[\left(G f G^T\right)\odot\left(B^T x B\right)\right]A
\end{align}
\clarity{where $G\in\RealNum^{(m+2){\times}3}$, $B^T\in\RealNum^{(m+2){\times}(m+2)}$, and $A^T\in\RealNum^{m{\times}(m+2)}$ are called transformation matrices.
Specifically, the $G$ and $B^T$ matrices transform the weights and the input feature maps, respectively, from the spatial domain to the Winograd domain.
Here, the convolution becomes an $(m+2)^2$-sized element-wise multiplication of the feature maps with the filter weights such that the number of multiplications is reduced from $m^2\cdot9$ to $(m+2)^2$.
Then, the $A^T$ matrix transforms the output feature maps back to the spatial domain.}


While larger feature map tile sizes ($m$ $\uparrow$) reduce the number of required multiply-and-accumulate operations (MACs $\downarrow$) compared to the standard convolution algorithm, this comes at the cost of more complex transformation matrices, higher sensitivity to numerical inaccuracies, and \clarity{so, in practice,} diminishing returns~\cite{barabasz2020error}.
Thus, the focus of actual implementations has been mainly put towards $m\in\{2,4\}$, resulting in a \clarity{potential} reduction of the number of \glspl{MAC} by 2.25$\times$ for $F_2$ and by 4$\times$ for $F_4$. 
Unfortunately, the numerical instability of $F_4$ prevents a straightforward adoption of \verb|int8| operations~\cite{Lavin2015a,fernandez2020searching,barabasz_2019_winogradbeyondlinear}.
Moreover, the challenges of processing more complex transformation matrices in a programmable AI accelerator have not been addressed in previous works~\cite{liu2021winocnn, lu2018spwa, yang2021biswsrbs}. 

This work aims primarily at enabling \verb|int8| Winograd $F_4$ inference on a domain-specific accelerator.
\clarity{Particularly,} we propose a novel tap-wise quantization algorithm to overcome the numerical issue of Winograd $F_4$ and an architectural and micro-architectural design space exploration for efficient hardware implementation. 
\clarity{An extensive evaluation demonstrates that the tap-wise quantization algorithm guarantees negligible accuracy drop on several state-of-the-art \glspl{CNN}.
Moreover, the domain-specific accelerator enhanced to support Winograd $F_4$ with tap-wise quantization runs up to \revised{3.42\x{}} faster on compute-intensive convolutional layers, and up to \revised{1.83\x{}} faster on an entire network with a gain up to \revised{1.85\x{}} in energy efficiency.}

In the following section, we introduce \clarity{the Winograd $F_4$ algorithm} in more detail, highlighting its challenges and our proposed solutions.

\section{Beyond Winograd 2×2}

The transformation matrices for Winograd \clarity{$F_2$} can be derived from the polynomial roots $\{0,1,-1\}$ s.t.: 

\smallEquationFont
\begin{align}
    B^T &= \begin{bmatrix}
1 & \phantom{-}0  & -1 & \phantom{-}0  \\
0 & \phantom{-}1  & \phantom{-}1  & \phantom{-}0  \\
0 & -1 & \phantom{-}1  & \phantom{-}0  \\
0 & \phantom{-}1  & \phantom{-}0  & -1
\end{bmatrix} &
G=\frac{1}{2}\begin{bmatrix}
2           & \phantom{-}0            & 0           \\
1 & \phantom{-}1  & 1 \\
1 & -1 &  1          \\
0           & \phantom{-}0            & 2          
\end{bmatrix} \nonumber \end{align}

\begin{equation}
A^T=\begin{bmatrix}
 1 & 1       & 1    & 0 \\
 0&1&-1&-1     
\end{bmatrix} \nonumber
\end{equation}

\normalsize
The matrices are relatively sparse and contain only $\pm1$ in $B^T$ and $A^T$, requiring \clarity{just} additions and subtractions. The weight transformation matrix has mostly $\pm\frac{1}{2}$ values, which can be implemented as a shift-by-1-and-add: $c=\frac{1}{2}a+\frac{1}{2}b=(a+b)$\texttt{>}\texttt{>}$1$.

To guarantee bit-true computation in the Winograd domain, $B^TxB$ requires only 2 extra bits ($k=2^2$), and $GfG^T$ requires 3 extra bits ($k=3^2$) as the sum of $k$ \verb|int|$n$ values results in a $\lceil\text{log}_2 (k(2^n-1)+1)\rceil$-bit integer in the worst case.
However, as weights and activation value distributions in \glspl{CNN} usually follow a Gaussian distribution centered around zero, in practice, 8 bits are sufficient to keep the same accuracy.

The most common Winograd $F_4$ algorithm uses the root points $\{0,1,-1,\frac{1}{2},\frac{1}{2}\}$, s.t.:
\scriptsize
\begin{align}
    B^T = \begin{bmatrix}
                     4 &  \phantom{-}0 &  -5           &  \phantom{-}0 &   1 &   0 \\
                     0 &  -4           &  -4           &  \phantom{-}1 &   1 &   0 \\
                     0 &  \phantom{-}4 &  -4           &  -1 &   1 &   0 \\
                     0 &  -2           &  -1           &  \phantom{-}2 &   1 &   0 \\
                     0 &  \phantom{-}2 &  -1           &  -2 &   1 &   0 \\
                     0 &  \phantom{-}4 &  \phantom{-}0 &  -5 &   0 &   1
\end{bmatrix},\quad
G=\frac{1}{3}\begin{bmatrix}
                   \phantom{-}3/4 &  \phantom{-}0 &  \phantom{-}0 \\
                   -1/2 &  -1/2 &  -1/2 \\
                   -1/2 &  \phantom{-}1/2 &  -1/2 \\
                   \phantom{-}1/8 &  \phantom{-}1/4 &  \phantom{-}1/2 \\
                   \phantom{-}1/8 &  -1/4 &  \phantom{-}1/2 \\
                   \phantom{-}0 &  \phantom{-}0 &  \phantom{-}3      
\end{bmatrix} \nonumber  \end{align}\ifdefined\LONG\else\vspace{-2mm}\fi
\begin{align}
A^T=\begin{bmatrix}
1 & 1 & \phantom{-}1 & 1 & \phantom{-}1 & 0 \\
0 & 1 & -1 & 2 & -2 & 0 \\
0 & 1 & \phantom{-}1 & 4 & \phantom{-}4 & 0 \\
0 & 1 & -1 & 8 & -8 & 1
\end{bmatrix} \nonumber 
\end{align}
\normalsize


The transformation matrices of $F_4$ are much less sparse, contain a wider range of coefficients, and require more computing effort. 
Therefore, while $F_4$ further reduces the number of MACs, it also introduces three significant challenges.

\begin{figure}
    \ifdefined\LONG
    \includegraphics[width=\linewidth]{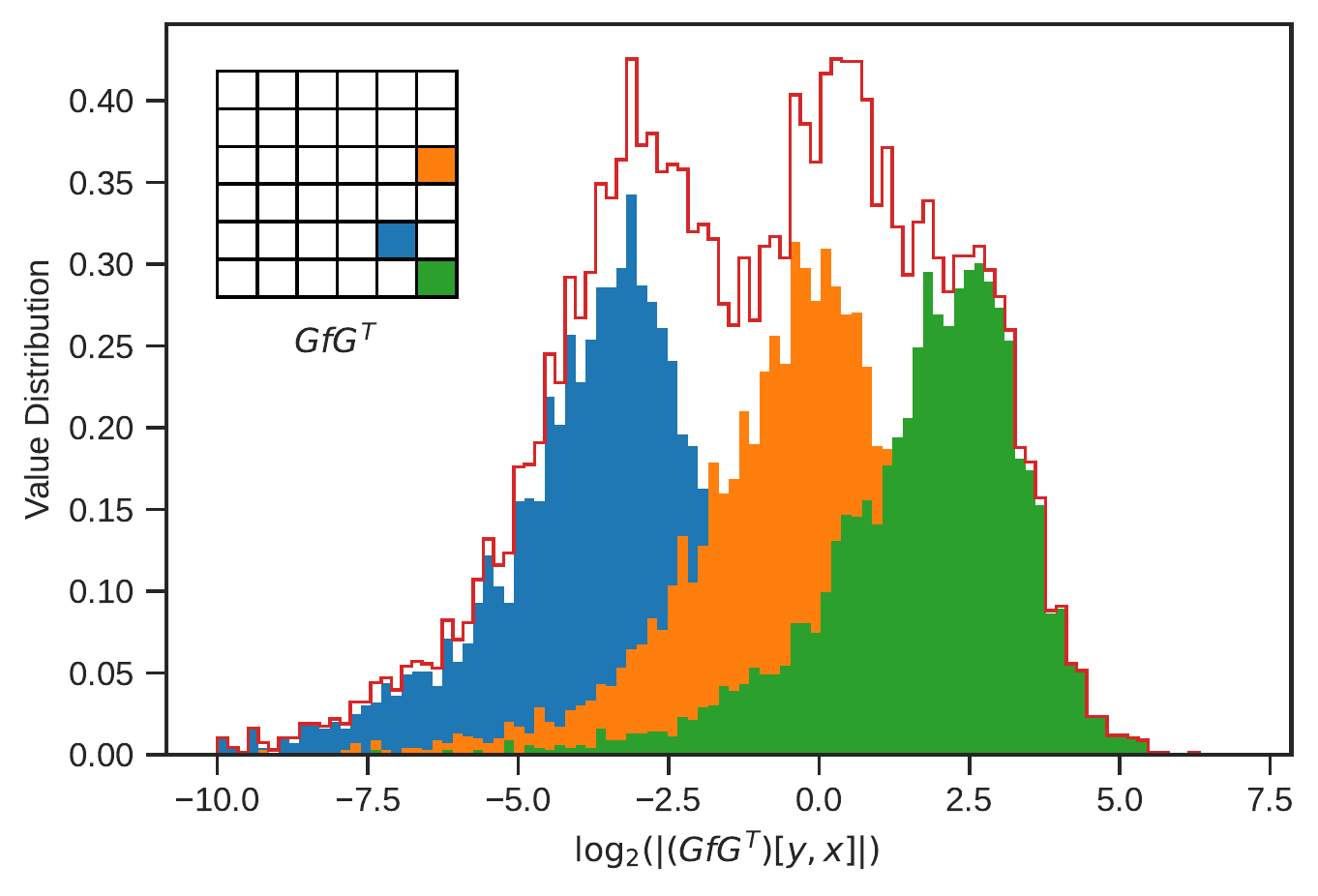}
    \else 
    \centering\includegraphics[width=0.8\linewidth,trim={0cm 0.3cm 0 0},clip]{figs/valuedistribution__GfG_rebuttal.pdf}\vspace{-2mm}
    \fi
    \caption{Weight Distribution in Winograd domain $GfG^T$ for 3 selected taps \revised{and their combined distribution }for ResNet-34 on ImageNet}
    \label{fig:GfGdistribution}
\end{figure}

\textbf{Challenge I: Non-uniform dynamic range.} 
A bit-true $F_4$ Winograd algorithm requires 10 extra bits for the weights and 8 extra bits for both the input and output feature maps transformations.
Clearly, such an increased bitwidth represents an unfeasible requirement for a high-throughput hardware accelerator as it significantly raises the power and area costs. 
However, quantizing all the taps to \verb|int8| in a traditional fashion, i.e., using the same scaling factor for all the taps, has a disruptive effect on the accuracy of the network~\cite{fernandez2020searching}.
We \clarity{found out} that the Winograd transformation matrices significantly change the dynamic range of each output tap, as shown in \cref{fig:GfGdistribution} for three taps of the weights in Winograd domain $GfG^T$ of ResNet-34. 

To this end, we propose a \textit{tap-wise quantization algorithm} to enable \verb|int8| inference with the $F_4$ Winograd algorithm.
Specifically, we present a training method that learns hardware-friendly powers-of-two scaling factors needed to independently quantize each tap based on its post-transformation dynamic range. 

\textbf{Challenge II: Complex transformation operations.} 
As the heart of virtually all modern accelerators is a large and high-throughput 2D or 3D data path for \gls{GEMM}~\cite{TPU2016, liao2019davinci, nvidiaa100}, translating the lower computational complexity of the Winograd algorithm into wall-clock time speed-up is not straightforward. 
Among all the steps involved in the Winograd algorithm, only the tap-wise multiplications can be efficiently processed as a batched \gls{GEMM}.
On the other hand, the input, output, and weight transformations involve several small \glspl{GEMM} and data-layout rearrangement operations, which cannot be processed at high throughput on a 2D or 3D \gls{GEMM} engine.
Thus, increasing the Winograd tile size moves ``ops'' from cheap, high-arithmetic intensity operations to more sparse, low-arithmetic intensity ones.

To address this challenge, we present a design space exploration of custom hardwired modules that implement the low-arithmetic intensity ``ops'' of the Winograd transformation operations in an area- and power-efficient way. 

\textbf{Challenge III: Orchestrating heterogeneous operations.}
One of the major challenges of developing a high-performance CNN accelerator is balancing memory bandwidth and compute throughput.
By adding a new class of operations, the Winograd algorithm increases the heterogeneity of the compute operations, making the orchestration of data movements and computations much more complex.   
Moreover, the Winograd algorithm lowers the computational complexity of the Conv2D operation compared to the \clarity{standard} implementation but at the cost of substantially reducing the data reuse opportunities.
This characteristic inevitably puts more pressure on the memory bandwidth requirements and calls for a careful dataflow development.

With this in mind, we show how to integrate the Winograd transformation engines in an industrial grade, programmable AI accelerator and how to \clarity{tune the microarchitecture of such blocks} to match the throughput of data movement, \clarity{Winograd} transformation, and compute operations, maximizing the overall compute efficiency.
\clarity{The presented methodology can serve as a guideline for DSA designers willing to exploit the Winograd transformation engines in other accelerators.}

%% file: 2_algorithm.tex
\section{Tap-Wise Quantization}

\textbf{Quantization.} Neural networks are commonly trained with floating-point numbers with 16--32\,bits. However, this comes with a significant power and area cost. Quantization to integer numbers for inference has become popular as most neural networks can be quantized to \verb|int8| with minimal or no accuracy loss \cite{zhu2020towards, guo2018survey}. Floating-point numbers are approximated as integer numbers with a shared FP32 scale factor $s$, s.t., $x_{\text{\texttt{float32}}}\approx\hat x_{\text{\texttt{int}}n}\cdots$
where $s = \frac{x_\text{max}}{2^{n-1}}$, and $x_\text{max}$ is the largest representable value, and the quantized value:
\begin{equation}
    \hat x_{\text{\texttt{int}}n}=\nint*{x/s}_{\text{\texttt{int}}n}=\text{clamp}\left(\nint*{x/s}, -2^{n-1}, 2^{n-1}-1\right)\label{eq:quant}
\end{equation}
We calibrate \clarity{$x_\text{max}$} by calculating a running average of the maximum values obtained during the training of the full network. After scaling, the data is rounded to the next integer value, and \clarity{clamped} within the $n$-bit integer number range, i.e., [-128,127] for \verb|int8|, denoted by the function $\nint*{x}_{\text{\texttt{int}}n}$. 



Previously, several works have proposed to quantize directly in the Winograd domain\cite{JiongGong2018,li2020lance,meng2019efficient,fernandez2020searching,barabasz2020quantaized}. Even though this helped to improve the performance of Winograd $F_2$ ($m=2$), it is is not sufficient for $F_4$ and larger tile sizes. Specifically, looking at the value distributions, we \clarity{found out} that the weights and feature maps in the Winograd domain heavily depend on their tap index, as shown in \cref{fig:GfGdistribution}. For this reason, we propose to independently quantize each tap.

\textbf{Tap-wise Quantization.} Based on the formulas of Winograd (\cref{eq:winograd}), and the general quantization (\cref{eq:quant}), \shepherd{tap-wisely} quantized Winograd (for a single \gls{oFM} and single tile) can be described as follows:

\begin{adjustbox}{width=\linewidth,center}
\begin{minipage}{\linewidth}
\vspace{4mm}
\begin{align}
y&=\sum_{C_{in}} A^T\left( s_B\nint*{B^T \hat x_{C_{in}} B/s_B}_{{\text{\texttt{int}}b}} \odot s_G\nint*{G \hat f_{C_{in}}G^T/s_G}_{{\text{\texttt{int}}b}}\right)A \nonumber  
\end{align}
\vspace{0.01mm}
\end{minipage}
\end{adjustbox}
We then replace the linear scaling factors $s_B$ and $s_G$ with tap-wise scaling matrices $S_G, S_B \in \RealNum^{(m+r-1)\times(m+r-1)}$ for the weights and the \acrfull{iFM} respectively. We define $S_{BG}=S_G\odot S_B$ for the \acrfull{oFM}. The multiplications/divisions with scalars are substituted with their element-wise counterparts $\odot, \oslash$.

Finally, we apply the distributivity law and reararrange the linear operations to obtain the following quantization scheme:

\begin{adjustbox}{width=\linewidth,center}
\begin{minipage}{\linewidth}
\vspace{2mm}
\begin{align*}
    A^T\!\left(\underbrace{S_B\!\odot\!S_G}_{S_{BG}}\!\odot\!\!\sum_{C_{in}} \underbrace{ \nint*{B^T \hat x_{C_{in}} B\!\oslash\!S_B}_{{\text{\texttt{int}}b}} \odot \nint*{G \hat f_{C_{in}}G^T\!\oslash\!S_G}_{{\text{\texttt{int}}b}}}_{{\text{\texttt{int}}2b}}\right)\!A.
\end{align*}
\vspace{0.01mm}
\end{minipage}
\end{adjustbox}

The multiplications and accumulations over the \glspl{iFM} are calculated in the integer domain, and rescaling is applied just once before the back-transformation, as an element-wise multiplication with ${S_{BG}}$.

\subsection{Winograd-aware Training}\label{sec:winogradaware}
We use stochastic gradient descent to train the network. To improve the training outcome, we also adopt the static Winograd-aware training method \cite{fernandez2020searching}, \clarity{which propagates the gradients through the Winograd domain}.

Fernandez et al.~\cite{fernandez2020searching} also proposed the \texttt{flex} variant, in which they learn the transformation matrices as a network parameter, and thus propagate the gradients back to $G, B, \text{and } A$. We are not considering this option because it introduces significantly more floating-point operations by making the transformation matrices dense and by preventing the use of HW-friendly shift-and-add operations.
\begin{figure*}[t]
  \centering
  \includegraphics[width=.95\textwidth]{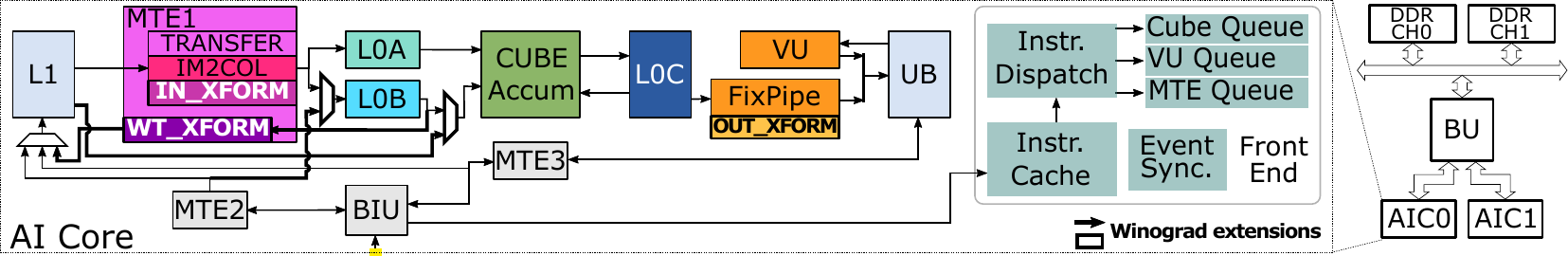}
  \caption{High-level overview of the inference accelerator with the proposed extensions.}
  \label{fig:system_overview}
\end{figure*}

\subsection{Power-of-two Tap-wise Quantization}\label{sec:power2}
The scaling operations cannot be moved outside of the Winograd domain and therefore introduce one multiplication with an FP32 value per transformation and tap. These multiplications can be shared among the output channels for \glspl{iFM} transformation and input channels for the \glspl{oFM} transformation. Nevertheless, it is favorable to restrict the scaling values to power-of-two values, such that the transformations can be performed with shift-and-add operations only, including the rescaling to adapt to the dynamic range. We evaluate and combine 3 approaches to learn the \clarity{power-of-two} scaling factors.

\textbf{Straight-forward power-of-two quantization.}
All scaling factors calculated from the calibrated maximum values are rounded to the next power-of-two: $\tilde s_{i,j} := 2^{\lceil{\log_2 s_{i,j}\rceil}}$, such that the quantized value is 
$q_{\text{\texttt{int}}b}(x) :=  \nint*{x / 2^{\lceil{\log_2 s\rceil}}}_{\text{\texttt{int}}b}$.

\textbf{Learned power-of-two quantization.} \clarity{The scaling factor can be learned to find a more optimal representation range, particularly as improving the precision of smaller values might be more important for the end-to-end accuracy compared to having less clamped values}.

The quantization function is a multi-step function and its derivative is zero almost everywhere thus preventing meaningful training. We approximate the gradient with the straight-through estimator~\cite{bengio2013estimating}: $\frac{\partial}{\partial x} \lceil{x}\rceil=\frac{\partial}{\partial x} \lceil{x}\rfloor=\frac{\partial}{\partial x} \lfloor{x}\rfloor=1$. Instead of training the scale value $s$, we calculate the gradient to the logarithm of $\log_2 t$, s.t.,  $s=2^{\lceil{\log_2 t\rceil}}$ \cite{jain2019trained}.

\begin{align}
    \frac{\partial q(x)}{\partial \log_2(t) }=s\;\text{ln}(2)\cdot\text{clamp}\left(\nint*{\frac{x}{s}}-\frac{x}{s}, -2^{b-1}, 2^{b-1}-1\right)
\end{align}
Furthermore, gradients need to be normalized for better convergence and to guarantee scale invariance for data and scaling factors, otherwise, they depend heavily on the relative distance of the input to the clamping threshold of the quantization function. For the scaling factors, we are using the Adam optimizer with its built-in gradient normalization ($\beta_1 =0.9$, $\beta_2=0.99$) \cite{kingma2014adam}. For the other parameters, we stick to standard SGD with a separate learning rate.

\textbf{Knowledge distillation.} Knowledge distillation (KD) has been used to train compact networks (student) \clarity{by minimizing the distance to a} larger network (teacher). We also use KD in the proposed training flow by setting the power-of-two tap-wise quantized network as the student and the floating-point baseline model as the teacher. We adopt the Kullback-Leibler divergence loss and the tempered softmax activation function \cite{hinton2015distilling}.

\revised{As not all convolutional layers can be implemented efficiently with the Winograd algorithm, we only replace those with a $3{\times}3$ kernel size and unitary stride, whereas all the others, like the $1{\times}1$ pointwise convolutions, are processed using a standard algorithm.} 
Although strided convolution can be implemented with the Winograd algorithm~\cite{yepez2020stride, yang2020stride}, the control and compute overhead dominates the potential \glspl{MAC} reduction (i.e., stride-2 $F_4$ \clarity{leads only to} a 1.8$\times$ \glspl{MAC} reduction). 

%% file: 3_hardwareAccel.tex
\section{Hardware Acceleration}
\label{sec:hardware_acceleration}\label{sec:hardware_acc}
\subsection{Baseline Accelerator}
\label{sec:baseline_acc}
\Cref{fig:system_overview} shows the architecture of our baseline inference DSA, featuring two AI cores (AIC0 and AIC1) inspired by the DaVinci architecture~\cite{liao2019davinci}.
Each core exposes a custom instruction set architecture (ISA) and implements all functionalities necessary for processing CNN layers.

The datapath of the AI core comprises a \unit{Cube Unit} for \glspl{GEMM}, a \unit{Vector Unit} for vector operations, and a \unit{Scalar Unit} to handle scalar tasks.
The \unit{Cube Unit} performs a \gls{GEMM} between two \verb|int8| matrices of size $[16{\times}32]$ and $[32{\times}16]$ to produce an \verb|int32| $[16{\times}16]$ output matrix, which can be optionally accumulated to a third input operand.
The memory access patterns are simplified by storing the input and output tensors for the \unit{Cube Unit} in the fractal format~\cite{optimizing_cnn_model_cpus_usenix19}, where the dimension of the tensor used as the reduction axis ($C$) is split into a sub-dimension of size $C_0=32$ and a super-dimension of size $C_1=\frac{C}{32}$.
Thus, for instance, the data layout of the \glspl{iFM} for a convolutional operation is \tensorDim{N}{C_1}{H}{W}{C_0}

The \unit{Vector Unit} is 256B wide and comprises multiple parallel lanes with a peak throughput of 128 FP16 or 256 \verb|int8| operations per cycle.
It performs general arithmetic operations between vectors besides more specific ones needed in CNN workloads, e.g., data type conversion, ReLU, and pairwise reductions.
The number of parallel lanes ensures that the throughput of the \unit{Vector Unit} matches the output data rate of the \unit{Cube Unit} for relevant CNN workloads.

The on-chip memory hierarchy follows a multi-level organization, where \unit{L0A} and \unit{L0B} serve as the input buffers for the \unit{Cube Unit} and \unit{L0C} as its output buffer, \unit{UB} as the input and output buffer for the \unit{Vector Unit}, and \unit{L1} as the second level memory.
The memory hierarchy is fully software-managed by the memory transfer engines (MTEs), which perform data movements and layout transformations.
Specifically, the \unit{MTE2} is in charge of transferring chunks of data from global memory (\unit{GM}) to \unit{L1} or \unit{UB}, whereas the \unit{MTE3} of transferring data from \unit{UB} to \unit{GM} or to \unit{L1} for fusing multiple consecutive layers.
The \unit{MTE1} transfers input tiles from \unit{L1} to \unit{L0A} or \unit{L0B} and can optionally perform the im2col transformation~\cite{im2col_paper} to lower a 2D convolution into a \gls{GEMM}.
The \unit{im2col} engine supports $3$, $5$, and $7$ as kernel sizes and $1$ and $2$ as stride parameters.
The \unit{FixPipe} module within the \unit{Vector Unit} transfers the output of the \unit{Cube Unit} from \unit{L0C} to \unit{UB}, potentially performing re-quantization operations on-the-fly.

The size and the number of banks of the on-chip memories are tuned to minimize the area while having enough bandwidth and capacity to avoid blocking the computational units.
Specifically, \unit{L0A} and \unit{L0B} can feed one operand per cycle to the \unit{Cube Unit} without incurring bank conflicts.
Similarly, \unit{L0C} can sustain write and read operations from the Cube Unit at the potential rate of one output tile per cycle, and it also has an additional read port towards the \unit{FixPipe} module.
\unit{L1} has a rather complex addressing scheme and multiple read and write ports, managing bank conflicts at run time.
The idle cycles caused by bank conflicts can be excluded from the critical path by exploiting data reuse in other memories.

The AI core relies on an in-order scalar front-end to offload instructions to the \unit{MTEs}, the \unit{Vector Unit}, and the \unit{Cube Unit}.
All units have a private instruction queue and a \textmu-sequencer to repeat the same instruction on different data and reduce the dispatching overhead.
Specifically, the current ISA of the core requires an instruction repetition factor and one stride parameter for each operand.
Furthermore, the core also implements a form of \emph{decoupled access/execute strategy}~\cite{dae_arch} as the different units are synchronized with an explicit token exchanging mechanism~\cite{vta_tvm}.
Such a mechanism allows the programmer to control the overlap between data movements and compute operations.

\revised{We decided to use an accelerator based on a \gls{GEMM} engine as our baseline since it represents a versatile and flexible design choice compared to a fully spatial architecture like Eyeriss~\cite{Chen2016} or MAERI~\cite{maeri}. 
Specifically, having an AI core grounded in linear algebra eases the development of the compiler infrastructure needed to support the growing diversity of AI workloads~\cite{norrie2021tpuv2v3}.
However, in the following section, we will present the microarchitectural space of the Winograd transformation blocks such that they can be tuned for the characteristics of the target accelerator system.}

\begin{figure*}[t]
    \centering
    \includegraphics[width=\textwidth]{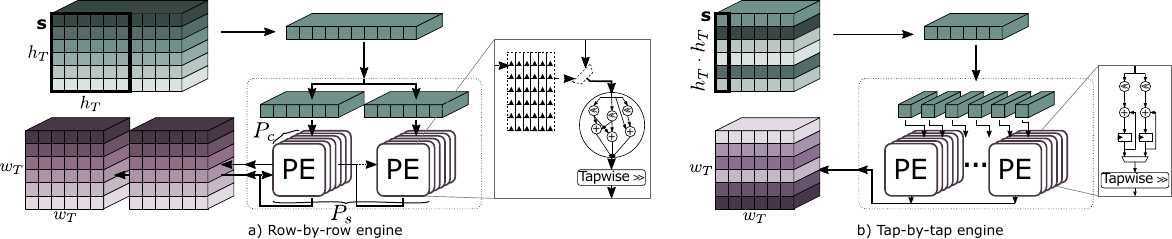}
    \caption{Winograd Transformation Engines.}
    \label{fig:wino_xform_engines}
\end{figure*}

\subsection{Winograd Extensions}
\label{sec:wino_extensions}
\subsubsection{Winograd Transformation Engines}
\label{sec:xform_engine}

The Winograd transformations $B^TxB$, $GfG^T$, and $A^TYA$ from \cref{eq:winograd} can be generalized as follows:
\begin{equation}
    s_w = T^T \times s \times T = T^T \times \Tilde{s},
    \label{eq:general_wino_xform}
\end{equation}
where $T$ is a generic constant $[h_T \times w_T]$ transformation matrix, $s$ is a $[h_T \times h_T]$ tile of the \glspl{iFM}, \glspl{oFM}, or weights, depending on the specific transformation.
The operations in \cref{eq:general_wino_xform} are repeated multiple times to transform the entire input tensor, namely, $\frac{H}{m}{\times}\frac{W}{m}{\times}C_{in/out}$ for the input/output transformations, and $C_{in}{\times}C_{out}$ for the weight transformation.

To design an area- and power-efficient transformation engine, we unroll the whole Winograd transformation (\cref{eq:general_wino_xform}) into a flat data flow graph (DFG), which can be heavily optimized by exploiting that we only use integer operations and that the $T$ matrix is constant and known at design time.
Specifically, as the transformation matrices have many symmetries and common terms, we apply common subexpression elimination (CSE) to share computations in time or space, reducing the number of cycles or resources needed.
\clarity{Moreover,} as most values in the transformation matrices are powers-of-two, we avoid using multipliers and carry out the computation using only shifters and adders.
The few multiplications with non-power-of-two numbers are split into multiple shift-and-add operations, e.g., $c=5\cdot a=(a<<2)+a$. 
The bitwidth is kept to the minimum for each intermediate operation.
Finally, we perform the scheduling and resource allocation of the DFG, exploring different area-throughput trade-offs, and selecting the solution that works best based on the requirements of the target transformation.
Specifically, we devise two high-level implementation strategies, which we denote as \textit{row-by-row} and \textit{tap-by-tap} transformation engines.

In the \textit{row-by-row transformation engine} (\cref{fig:wino_xform_engines}a), we decompose the transformation operation (\cref{eq:general_wino_xform}) as a series of vector-matrix multiplications $s[y,:] \times T$, which we map on a spatial processing element (PE).
Specifically, the PE reads one row at a time of the matrix $s$ and performs the entire operation $\Tilde{s} = s \times T$ in multiple cycles.
The PE hardcodes the multiplication of the input vector with the matrix $T$, using only adders and fixed shifters.
The second part of the Winograd transformation, namely, $T^T \times \tilde s$, can be computed using the already allocated resources (\textit{slow solution}) or using additional spatial resources inside the PE (\textit{fast solution}).
The former saves resources at the cost of higher latency, requiring only an additional set of $h_T{\cdot}w_T$ registers to store the intermediate results.
The latter requires additional $w_T\cdot w_T$ lanes to compute $s_w$ in an output-stationary fashion, reducing the number of required cycles but increasing the number of adders needed.
Moreover, the former solution produces one row of the output matrix at a time, whereas the latter produces the entire output matrix.
As shown in \cref{fig:wino_xform_engines}a, the PE can be replicated to perform multiple transformations in parallel.
$P_c$ and $P_s$ represent the two factors controlling the number of parallel transformations along the channels and the spatial dimension, \clarity{respectively}.
The parallelization strategy is not only constrained by the area \clarity{budget} but also by the memory bandwidth and access pattern requirements.
Specifically, the row-by-row engine requires all the elements of a spatial row to be contiguous in memory and the memory bandwidth to be sufficient for reading multiple rows of different tiles in the spatial dimension.
Similar considerations also apply to the input channels dimension.

The \textit{tap-by-tap transformation engine} (\cref{fig:wino_xform_engines}b) represents a different point of the optimization space, where the DFG of \cref{eq:general_wino_xform} is completely unrolled in time.
The PE is very simple in this solution, comprising only a configurable shifter, an adder/subtractor, and an accumulator register.
Thus, in the worst case, the PE takes $h_T {\cdot} h_T$ cycles to compute one tap.
Luckily, we can exploit two properties of the transformation matrices to reduce the total number of cycles:
i) the transformation matrices are sparse, lowering the average number of cycles needed per tap;
ii) some taps share a significant fraction of computations with other taps, so we can apply CSE in time to avoid recomputations.
Higher throughput can be achieved by replicating the PEs to perform multiple transformations in parallel.
Apart from $P_c$ and $P_s$, we can use the number of parallel taps in a single PE, $P_t$, as an additional parallelization axis.
Since one input value can be read once and shared to compute multiple taps in parallel, increasing $P_t$ does not affect the input bandwidth requirements.
Moreover, by splitting the write back into multiple sub-writes, $P_t$ does not affect the output bandwidth requirements either.

\input{tables/engines.tex}

We enhance the PE with an input or an output stage comprising a configurable shifter and a rounding module to support tap-wise quantization. The number of quantization stages depends on the number of taps produced or consumed in parallel per cycle. 
The overall performance and requirements for the two implementation styles are summarized in \cref{tab:engines_requirements}.

In the following section, we will detail the data flow of the Winograd Operator, illustrating the motivations behind our specific design choices.

\subsubsection{Winograd Convolutional Operator}
\label{sec:winograd_operator}
The baseline is extended with the transformation engines (reported in bold in \cref{fig:system_overview}) needed for the \gls{iFM} and weight transformations in the \unit{MTE1}, and for the \gls{oFM} transformation in the \unit{FixPipe} module. 
The data flow of the Winograd operator for a $3{\times}3$ Conv2D layer is reported in \cref{lst:wino_operator}.
It refers to the computation of a subset of the channels of the \glspl{oFM}, and it works as follows.

First, a tile of the weights is transferred from \unit{GM} to \unit{L1} and transformed to the Winograd domain (lines~2 -- 6).
Each core operates on different output channels of the weights (line~2).
\unit{L0B} is used as an intermediate buffer to store the weights before the transformations.
Thus, the data transfer is carried out in tiles (line~5), each tile is transformed using the weight transformation engine within the \unit{MTE1} module, and the output is stored in \unit{L1} (line~6).
In order to overlap data transfers and the weight transformation, \unit{L0B} is double-buffered (line~4), and proper token exchanging instructions synchronize the transfers performed by the \unit{MTE2} with the transformations performed by the \unit{MTE1}.
For simplicity, such synchronization instructions are not shown in the pseudocode.
The transformed weights are kept stationary in \unit{L1} and reused for all the \glspl{iFM}.
Three levels of loop blocking are used to perform double-buffering across the entire on-chip memory hierarchy, maximizing the concurrency between compute, data movement, and \clarity{Winograd} transformation operations.
Specifically, in the outer-most loop block (lines~8 -- 10), the load of the \glspl{iFM} from \unit{GM} (line~11) is overlapped with all the core-level compute and data-movement operations (lines~13 -- 26).
In the loop block in the middle (lines~13 -- 15), the input transformation and the cube operations (lines~17 -- 23) are done in parallel with the output transformation (line~24), the re-quantization step (line~25), and the write to \unit{GM} (line~26).
In the innermost loop block (lines~17 -- 20), the input transformation (line~21) is overlapped with the batched \glspl{GEMM} performed in the \unit{Cube Unit} (lines~22 -- 23).

\input{algos/operator}

Matching the production and consumption rates is key to maximizing the \unit{Cube Unit} utilization, so compute efficiency.
As any overhead in the innermost loop will be multiplied by the total number of outer iterations, matching the input transformation production rate with the Cube Unit consumption rate has the highest priority.
As reported in \cref{sec:baseline_acc}, the fractal data layout \tensorDim{N}{C_1}{H}{W}{32} is used for the \glspl{iFM} in \unit{L1}, making 32 input channels and the spatial dimension $W$ contiguous in memory.
This represents a perfect fit for the row-by-row engine as, given the read bandwidth of \unit{L1} and the write bandwidth of \unit{L0A}, it allows us to replicate the PEs 32 times along the $C_{in}$ dimension and two times along the $\frac{W}{4}$ spatial dimension, performing up to 64 transformations in parallel.
With this parallelism, the transformation engine has a production rate of $64 {\cdot} \frac{36}{12} \SI{}{B/cycle}$, which is 4\x{} slower than the consumption rate of the \unit{Cube Unit}.
Thus, to avoid blocking the \unit{Cube Unit}, we need to reuse the transformed \glspl{iFM} four times across the output channels, with two main consequences.
First, we need to compute at least $4{\cdot} 16$ output channels at a time.
Second, \unit{L0C} should store at least $64{\cdot} 16 {\cdot} 36$ \verb|int32| elements, that is a total size of $288\,kB$ considering double-buffering.
Note that the row-by-row engine produces multiple taps per cycle, which are read by the \unit{Cube Unit} in different cycles.
Thus, we enhance the addressing capabilities of \unit{L0A} with a \textit{diagonal} write mode such that different rows of different banks can be accessed with one single memory access.
As reported in the experimental results section, this modification has a negligible area and power overhead.

To keep the overall pipeline busy, the computations related to an input tile must be overlapped with the output transformation of the previous tile.
In the case of the output transformation engine, the choice of the transformation engine to use was mainly driven by the need to minimize the number of memory accesses.
The tap-by-tap engine reads multiple times the same data from the input memory, which is too costly in the case of \unit{L0C} as data is stored in \verb|int32|.
Thus, we rely on the row-by-row engine for the output transformation engine.   
With the available \unit{L0C} bandwidth, up to $16$ transformations can be performed in parallel along the output channel dimension.
Thus, a volume of $36 \cdot C_{out} \cdot \frac{H}{4}\cdot\frac{W}{4}$ feature map (taps) in the Winograd domain will be transformed in \revised{$\frac{C_{out}}{16}\cdot\frac{H}{4}\cdot\frac{W}{4}\cdot10$} or $\frac{C_{out}}{16}\cdot\frac{H}{4}\cdot\frac{W}{4}\cdot6$ cycles, depending on the chosen row-by-row engine solution.
The same volume of data is produced by the Cube in $\frac{C_{out}}{16} {\cdot} \frac{Ci}{32} {\cdot} \frac{H}{4} {\cdot}\frac{W}{4} {\cdot} \frac{1}{16} {\cdot} 36$ cycles.
To match the two operations, we need at least $3$ fractal input channel tiles ($C_{in}{=}96$) for the fast engine and $6$ fractal input channel tiles for the slow one ($C_{in}{=}192$).
As many layers of SoA networks have less than $192$ input channels, we decide to use the fast engine.
Moreover, as the \unit{Cube Unit} writes the taps in different rows of \unit{L0C}, the output transformation engine performs a gather operation to collect one row of the input matrices.

Finally, the read and write operations to and from \unit{GM} must match the processing time of all the core-level operations.
As the weights are read from \unit{GM} and transformed on the fly (lines~2-6), the throughput of the weight transformation engine should match the external bandwidth. 
In this case, we rely on the tap-by-tap transformation engine as it produces the output data precisely in the data layout expected by the \unit{Cube Unit} for the weights.
Moreover, the layout of the weights can be reorganized offline such that we can avoid gathering operations when the PE reads the weights from the memory.
With two available AI cores, we need to read at least $2\cdot 18 \cdot 18 \cdot C_{in}$ B of \glspl{iFM} and to write $2\cdot 16 \cdot 16\cdot 64$ B of \glspl{oFM} to match the cores throughput, which corresponds to a \unit{GM} bandwidth of $\approx 2\cdot72 + \frac{7281}{C_{in}} \SI{}{B/cycle}$ assuming a peak compute efficiency for the core-level operations (lines 13-26).
Being this requirement hardly met when the AI Core clock frequency is in the order of the hundreds of \,MHz, we apply three system-level and data flow optimizations. 
First, as the two cores work on different sets of output channels, the \glspl{iFM} can be shared between the two cores, almost halving the required bandwidth. 
To this end, the cores are connected to the memory controllers via a \unit{Broadcast Unit} (\unit{BU} in Fig.~\ref{fig:system_overview}). 
The \unit{BU} can either accept independent memory requests from the \unit{MTEs} of the two cores and transfer them to the memory controllers or process special broadcast requests in the form of a streaming access pattern~\cite{streaming_nowatzki}. 
When the BU reveives two broadcast memory requests from the two cores, it acts as a DMA and brodcasts data from GM to the MTEs of the cores. 
To avoid deadlock, the \unit{BU} has two separate queues for non-broadcast and broadcast requests, where the latter are served with higher priority. 
Second, when the \glspl{iFM} shape is larger than $18{\times}18$, the volume to be transferred can be reduced by exploiting the halo regions that characterize the unitary-stride $3{\times}3$ Conv2d operator. 
Third, by prefetching input tiles and allocating multiple output buffers (instead of just two for double buffering), it is possible to decouple read and write operations, prioritizing the more critical read transfers.

%% file: tables/engines.tex
\begin{table}[t]
\centering
\footnotesize
    \renewcommand{\arraystretch}{1.25}
    
    \ifdefined\tableninja \small \setlength{\tabcolsep}{5pt}\else \setlength{\tabcolsep}{4pt} Switch \tabularxtabularx \fi    
    \caption{Performance and bandwidth requirements of the Winograd transformation engines. \label{tab:engines_requirements}}
\footnotesize\begin{tabular}{lrrrr}
\toprule
 & 
 \textbf{\begin{tabular}[r]{@{}r@{}}Cycles/Xform \\$[\SI{}{cycles}]$\end{tabular}} & 
 \textbf{\begin{tabular}[r]{@{}r@{}}Parallel\\Xforms\end{tabular}} & 
 \textbf{\begin{tabular}[r]{@{}r@{}}RD BW\\ $[\SI{}{B/cycle}]$\end{tabular}} &
 \textbf{\begin{tabular}[r]{@{}r@{}}WR BW\\$[\SI{}{B/cycle}]$\end{tabular}}
 \\
\hline

\multirow{1}{*}{Row-by-row} &
&
&
&
\\

\textit{\ - slow} &
$h_T + w_T$ &
\multirow{2}{*}{$P_c \cdot P_s$} &
\multirow{2}{*}{$P_c \cdot P_s \cdot h_T$} &
$P_c \cdot P_s \cdot h_T$ \\
                     
\textit{\ - fast} &
$h_T$  &
&
& 
$P_c \cdot P_s \cdot w_T \cdot w_T$  \\ \midrule

Tap-by-tap & 
$T$ dependent &
$P_c \cdot P_s \cdot P_t$ &
$P_c \cdot P_s$ &
$P_c \cdot P_s$     
\\ \bottomrule                     

\end{tabular}
\end{table}

%% file: algos/operator.tex
\begin{lstlisting}[language=python,
float,
label={lst:wino_operator},
caption={Dataflow of the Winograd operator},
style=mystyle]
# WEIGHT Blocking
for wt_tile_gm in WT_GM[2*cout_tile_sz + block_id:...].get_next_tile(
    cin_l0b_tile_sz, cout_l0b_tile_sz,
    double_buffering=True):
  mte2.transfer(wt_tile_gm, wt_tile_l0b)
  mte1.weight_xform(wt_tile_l0b, wt_tile_l1[...])
# IFM L2 Blocking
for ifm_gm_tile in IFM_GM.get_next_tile(
    batch_l2_tile_sz, h_out_l2_tile_sz,
    w_out_l2_tile_sz, double_buffering=True):
  mte2.transfer(ifm_gm_tile, ifm_l2_tile, broadcast=True)
  #IFM L1 Blocking
  for ifm_l1_tile in ifm_l2_tile.get_next_tile(
      batch_l1_tile_sz, h_out_l1_tile_sz,
      w_out_l1_tile_sz, double_buffering=True):
    # IFM L0 Blocking
    for ifm_l0_tile in ifm_l1_tile.get_next_tile(
        bath_l0a_tile_sz, h_out_l0a_tile_sz,
        w_out_l0a_tile_sz, chs_in_l0a_tile_sz,
        double_buffering=True):
      mte1.input_xform(ifm_l0_tile, ifm_l0a_tile)
      cube.mmad(ifm_l0a_tile, wt_tile_l1[...],
                out_l0c_tile[...])
    vec_unit.out_xform(out_l0c_tile, out_prescale_ub_tile)
    vec_unit.dconv(out_prescale_ub_tile, out_postscale_ub_tile, alpha_q)
    mte3.write(out_postscale_ub_tile, OUT_GM[:])
\end{lstlisting}

%% file: 4_ExperimantalEvaluation.tex
\section{Experimental Evaluation}

\subsection{Tap-wise Quantization Algorithm}

\subsubsection{Datasets and Baseline Networks}
We use two common image classification datasets, namely, CIFAR-10 \cite{krizhevsky2009learning} and ImageNet ILSVRC \cite{Russakovsky2014}, to compare \clarity{the proposed quantization flow} with other Winograd-based algorithms. CIFAR-10 has 60k 32\x 32 RGB images \clarity{divided into} 10 classes, and ImageNet 1.4M 224\x224 RGB images \clarity{divided into} 1k classes. We split the datasets into training (90\% of 50k/1.3M training set), validation (10\% of 50k/1.3M training set, used for learning rate scheduler), and test set (10k/100k, inference only). We use the standard preprocessing methods: random horizontal flip (training set only) and color normalization for CIFAR-10; resize, random crop, and color normalization for ImageNet. 

We benchmark ResNet-34 and ResNet-50 for the ImageNet dataset, where we use the pre-trained networks from Torchvision. The baseline networks (im2col/FP32) achieve 72.6\%/ 75.5\% Top-1 and 90.7\%/92.6\% Top-5 accuracy on the test set. For CIFAR-10, we re-implement the ResNet-20 \cite{He2015} and train it from scratch (94.4\%). Furthermore, we use a light-version of VGG \cite{Nagadomi2014}, used by Liu et al. \cite{liu2018efficientsparse} and Lance et al. \cite{li2020lance}, and replace all but the last dropout layers with batch normalization layers (92.2\%). We trained the networks using PyTorch while extending the Winograd-Aware Training \cite{fernandez2020searching} with tap-wise quantization support.

\subsubsection{Tap-Wise Quantization Evaluation}
\label{sec:tap-wise_quant_eval}
We retrain the network as a quantized \verb|int8| network from the FP32-baseline, whereas the weights and feature maps are quantized, as described in \cref{eq:quant}. All networks can be trained using 8-bit integers without any loss of precision. \clarity{We train the Winograd $F_2$ ResNet34 on Imagenet following the (static) Winograd-Aware algorithm (\cref{sec:winogradaware}), achieving} 71.4\% (-1.2\% drop) with 8-bit quantization. As expected, extending weights and feature map to 10 bits in the Winograd domain achieves the full accuracy because just 3\,bits are required for a bit-true calculation. Furthermore, we train \clarity{the Winograd $F_4$ version} with the Winograd-Aware method and \clarity{KD}. The baseline Winograd $F_4$ shows a significant drop of 13.6\%; even with two extra bits, the accuracy drops at least by 3.5\%.

\input{tables/winograd_quant_resnet34}

\Cref{tab:overviewapproaches} gives an overview of the performance of ResNet-34 on the ImageNet dataset, comparing several training methods and configurations. In the second section of the table, we evaluate the \clarity{tap-wise} quantization with unrestricted quantization scaling factors, i.e., $s_{i,j}\in \text{FP32}$. We further relax the numerical pressure by adding 2 extra bits in the Winograd domain, but keeping 8-bits in the spatial domain, denoted with \verb|int8/10|.

Using Winograd-aware static training and straight-forward threshold calibration, \clarity{the accuracy loss is small for full-}\verb|int8| (-1.2\%), \clarity{and the final accuracy is even closer} to the baseline (72.0\%, -0.6\%) with the 2\,bit extension in the Winograd domain. With knowledge distillation, \clarity{we can get the best performance even with} the full-\verb|int8| configuration.

As explained in \cref{sec:power2}, we prefer to have powers-of-2 quantization as all re-quantization and de-quantization operations become shifts within the Winograd domain. \clarity{The results with powers-of-2 scaling factors are summarized in the third section of the table. The straightforward method leads to} a 1.7/0.5\% drop with \verb|int8| and \verb|int8/10|. With knowledge distillation, the drop can be reduced to 0.4\% for the \verb|int8/10|. The best accuracy is achieved with $\log_2$ gradients and knowledge distillation, namely, 71.1\% (-1.5\%) for \verb|int8| and 72.3\% (-0.3\%) for \verb|int8/10|. \revised{The training with $\log_2$ gradients without knowledge distillation shows worse performance than the straightforward calibration method due to convergence issues. Knowledge distillation stabilizes the $\log_2$ training by acting as an implicit regularizer~\cite{saglietti2022solvable}.}

We further investigate the bit shifts learned by the network.  The feature maps are shifted right by 1 to 5 bits (i.e., $s \in \{2,2^2,\dots,2^5\}$) and the weights by 2 to 10 bits. Particularly, the large \clarity{difference} between the shift values implies why \clarity{quantizing with} a single scalar cannot work well. Within the same tap, the distribution is in the range of 2--3 bits, but it needs to be learned independently per layer.

\input{tables/winograd_soa}

\subsubsection{SoA Winograd-aware Quantization Methods}
\label{sec:qat_eval}

Recently, several approaches for Winograd-aware quantization have been presented. \Cref{tab:overviewSoA} gives a full overview of the main methods and of our solution. 
As the baseline accuracy varies across related works due to different training methods and implementation details, we \clarity{report their baseline accuracy and compare relative performance}. Our results are trained with the Winograd-aware method and powers-of-two tap-wise quantization, $\log_2$ gradients, and KD. The first network is ResNet-20 on CIFAR-10. Without adding any extra compute, we improved the (static) Winograd-Aware WA accuracy from 84.3\% to the full accuracy of 94.4\% with powers-of-two tap-wisely quantized $F_4$ \cite{fernandez2020searching}. 
Moreover, we also outperform both the WA-flex method, which trains the transformation matrices to recover some of the quantization losses \cite{fernandez2020searching}, and the Legendre-$F_4$ \cite{barabasz2020quantaized} method by 1.9\% and 2.1\%, respectively. 

We retrain the light version of VGG (VGG-nagadomi \cite{Nagadomi2014}). Our baseline network reaches 92.0\% accuracy, the tap-wisely powers-of-two quantized $F_4$ performs with a drop of 1.2\%, 0.1\% or no drop (for \verb|int8|, \verb|int8/9|, and \verb|int8/10|). For this network, no $F_4$ numbers have been presented in previous work, so we compare to $F_2$ results. Liu et al. \clarity{prunes the weights} in the Winograd domain to further reduce the compute intensity. They can retain their baseline performance of 93.3\% \cite{liu2018efficientsparse}. Li et al. proposed to quantize to \verb|int8| in the Winograd domain, where they can achieve 90.3\% (-0.1\%) \cite{li2020lance}.

Finally, we train and compare ResNet-50 on ImageNet. We achieve 75.2\%/92.3 (-0.3\%/-0.3) with 8 bits and 75.5\%/ 92.5 (0.0\%/-0.1\%) when extending the Winograd domain to 10 bits (\verb|int8/10|).\clarity{On such a benchmark, there are three main related} works. Meng et al. uses complex root points leading to numerically more stable but complex transformation matrices \cite{meng2019efficient} with a small drop, but with lower baseline (i.e., 73.2, -0.1\%). Liu et al. uses the Residue Number System RNS. They use very large tile size of (i.e., $14\times 14$) to compensate for the transformation overhead. Even though the RNS could perform the \verb|int8| operations losslessly (transformations and elementwise multiplications), the accuracy drops by 1.0\% or 75.1\%. It is expected that this is due to the quantization of the transformations matrices, furthermore, it is known \cite{alam2022winograd, Lavin2015a} that very large tiles introduce very high numerical error even for FP32. 
A very interesting approach is LoWino, they operate on FP32 weights and feature maps, but quantize linearly to 8\,bits before and after the elementwise multiplication in the Winograd domain \cite{li2021lowino}.
They achieve an identical accuracy of 75.5\% as our method with \verb|int8/10|, although with an accuracy drop of 0.6\% instead of 0\% as they start from a higher baseline. While LoWino provides the same reduction in operation count, it requires a 4\x{} higher bandwidth than our proposed method, which eliminates any benefits as shown in \cref{sec:throughputanalysis}. 
Notably, only our method with \verb|int8/10| can avoid any accuracy drop with ResNet-50. 
\begin{figure*}
     \centering
     \subfloat[Spatial Quantization]{
    \includegraphics[width=0.40\textwidth]{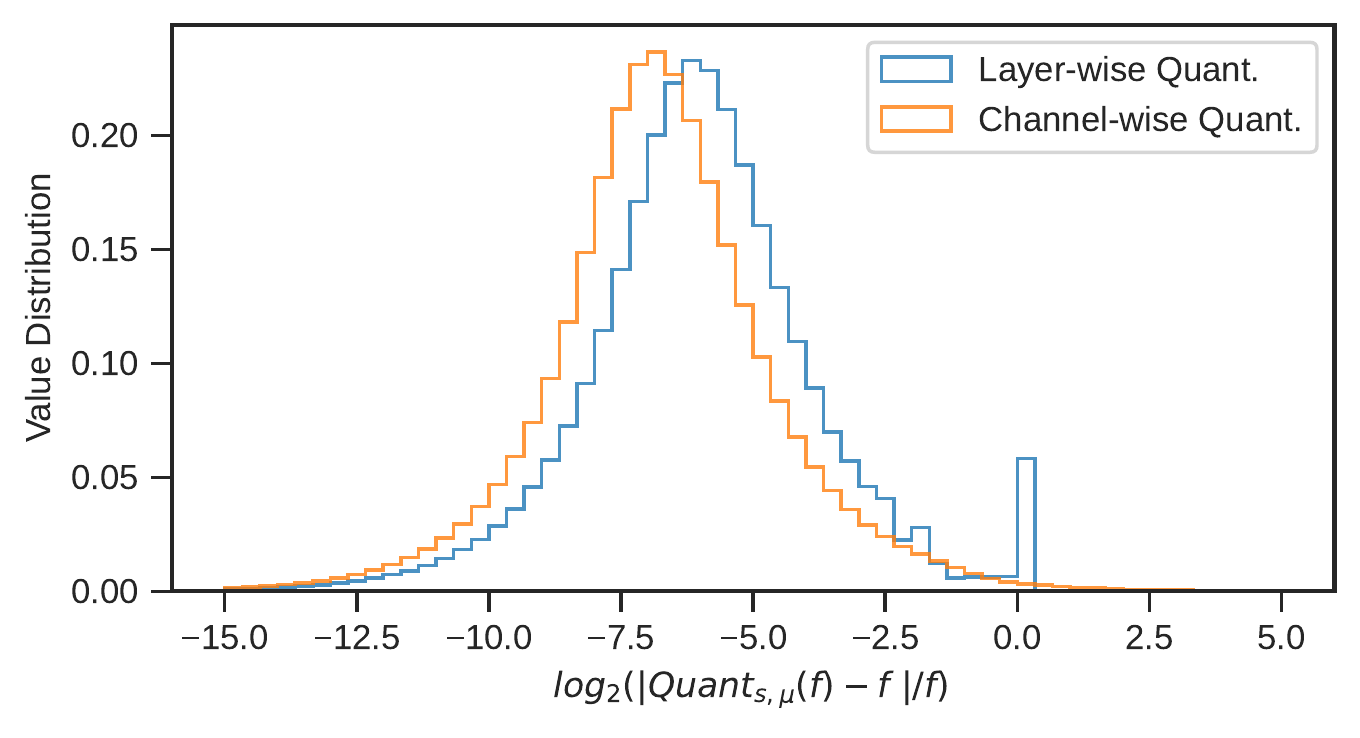}\label{fig:quanterror_spatial}

    }
    \qquad
    \subfloat[Quantization in Winograd domain]{
    \includegraphics[width=0.40\textwidth]{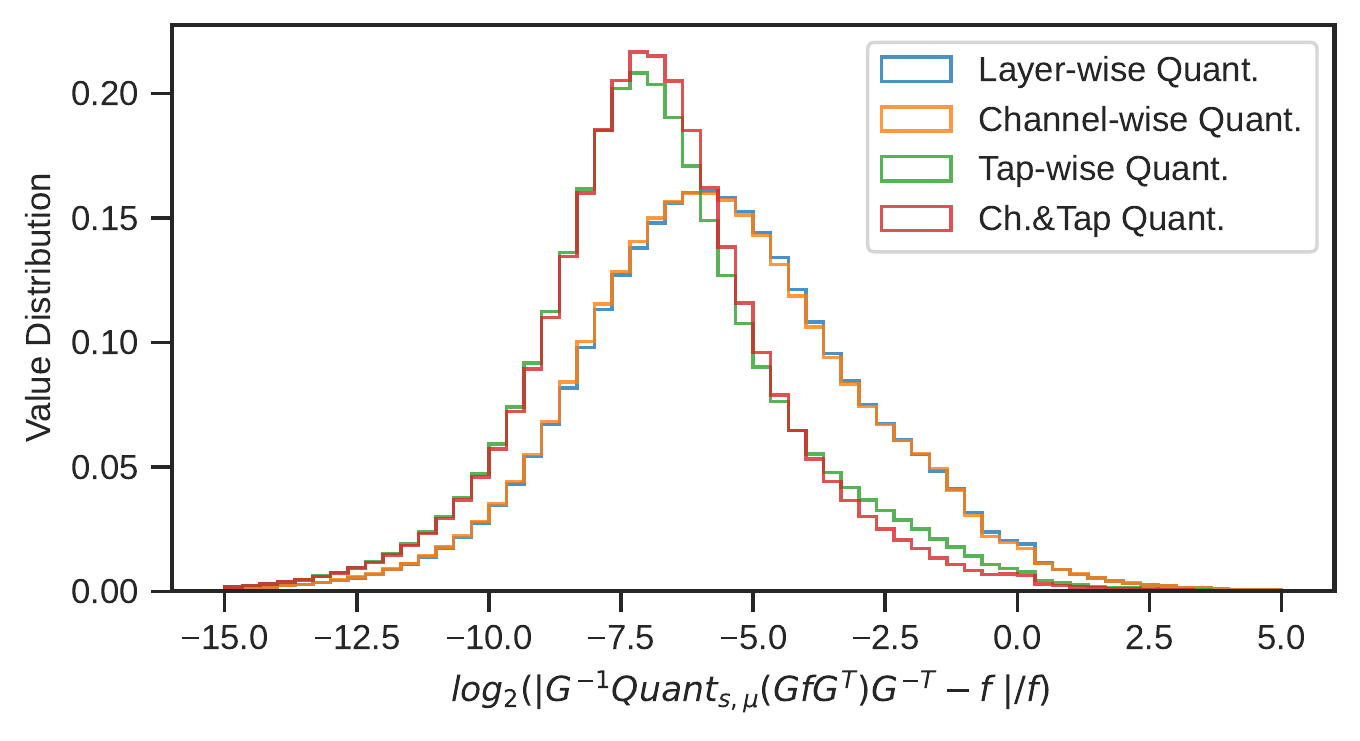}\label{fig:quanterror_wino}
    }
          \caption{\revised{Quantization error for the weights in (a) spatial and (b) Winograd domain on ResNet-34 using different strategies: layer-wise quantization, channel-wise quantization, tap-wise quantization, and  channel- \& tap-wise quantization.}}
\end{figure*}

\subsubsection{\revised{Tap-wise vs. Channel-wise Quantization}}\label{sec:finegrained}
\revised{Previous works have shown that fine-grain quantization strategies, particularly (output) channel-wise quantization, can significantly improve the accuracy of quantized networks \cite{krishnamoorthi2018quantizing,liang2021pruning}. 
Therefore, in this section, we compare the tap-wise with the channel-wise quantization strategy. 
We use a pre-trained ResNet-34 from Torchvision, and we evaluate the quantization error on the weights, although similar trends can also be observed for the feature maps.
The scaling factors $s$ are determined as follows:} 
\begin{align*}
    \hat{\gamma}=\arg\min_{\gamma} \sum_f |\mathrm{Quant}_{\mu,s}(f)-f|/|f|, \quad s=\gamma\sigma/2^{n-1},
\end{align*}

\revised{where $Quant_{s,\mu}(x) = \mu+s\nint*{(x-\mu)/s}_{\text{\texttt{int}}n}$, and the mean $\mu$, the standard deviation $\sigma$, and the optimized scaling factor $\hat{\gamma}$ are obtained per layer (uniform quantization strategy), per channel, or per tap.}

\revised{Fig.~\ref{fig:quanterror_spatial} shows the distribution of relative quantization error (in log2 scale) of all layers with kernel size $3\times 3$ for a uniform and for a channel-wise quantization strategy in the spatial domain for $n=8$ bits. 
Channel-wise quantization reduces the mean relative error by $1.7\times{}$, namely, from $2^{-6.01}$ to $2^{-6.72}$. 
Fig. \ref{fig:quanterror_wino} shows, instead, the error distribution for uniform, channel-wise, and tap-wise quantization in the Winograd domain.
Specifically, we quantize in the Winograd domain (i.e., $Quant(GfG^T)$), and then we transform the values back to the spatial domain to compare the error introduced by the quantization process. 
We calculate the Moore-Penrose inverse of the transformation matrices based on SVD to transform the quantized weights back to the spatial domain. 
In this case, the improvement of channel-wise quantization is significantly lower as the mean relative error reduces from $2^{-5.58}$ to $2^{-5.62}$. 
On the other hand, tap-wise quantization shows much better performance ($2.3\times$), leading to a mean error as low as $2^{-6.78}$. 
Combining channel-wise with tap-wise quantization further improves the average error by $1.06\times$ at the cost of a much more complicated compute phase. For networks with significantly different channel distribution, the combined quantization strategy might achieve better performance.}

\subsection{System Evaluation}
\label{sec:system_eval}
\subsubsection{Experimental Setup}
\label{sec:system_eval:setup}

\textbf{Area and Power.} To assess the area and power consumption of the accelerator, we developed the RTL of the parts of the AI core most affected by the Winograd extensions, specifically the \unit{Cube Unit}, the \unit{MTE1}, and the \unit{FixPipe} module.

{{We have implemented the design with a high-$k$ metal gate (HKMG) 28\,nm CMOS technology, and a corresponding multi-VT standard cell library at a supply voltage of 0.8\,V in typical corner. We have synthesized the design and performed place-and-route, targeting a clock frequency of $500$\,MHz in typical operating conditions. We used an industrial-grade memory compiler for the SRAM and register file macros.  
We have selected input data segments from the first 3\x{}3 layer of a ResNet-34 quantized using our method. Then we run a timing-annotated (SDF) post-place \& route gate-level simulations with a cycle-accurate event-based RTL simulator. Finally, we simulate the power consumption based on the extracted switching activities (VCD).}}

\textbf{Performance Profiling}.
An event-based simulator~\cite{villa2021need} was developed to model and profile the overall system.
Besides modeling timing behavior, the simulator also models data movements and computation to check the correctness of the results.
It was validated with micro benchmarking against the parts of the system developed in RTL.
\revised{Specifically, we compared the number of cycles obtained from the RTL simulation with that estimated by the simulator. On several small and medium-sized Conv2D operations, the simulator shows a $5\%$ worst-case difference.}
The simulator implements a \clarity{simple} model for the DRAM subsystem in which memory requests are served in order.
Moreover, the completion time of a memory request depends on the maximum bandwidth ($\SI{81.2}{B/cycle}$), which corresponds to \revised{$\approx0.8\cdot\SI{51.2}{GB/s}$ given the clock frequency of the core, and on a fixed average latency (150 AI core cycles) with a jitter extracted from a zero-mean Gaussian distribution with a variance of 5 cycles.
Such memory bandwidth and latency characteristics meet the expected performance of an LPDDR4x-3200 memory with two channels \cite{steiner2021exploration,hackenberg2015energy}}.
Although we do not use a detailed model of all the DRAM resources (e.g., command scheduling, channel bandwidth, bank, and row-buffer conflicts), their effects on the performance of the cases under analysis are minimal~\cite{villa2021need} as the memory accesses are regular and follow a streaming pattern.
We estimated the energy consumption by projecting the power consumption of computational units and memory obtained from the back-annotated gate-level simulations.
For \unit{L1}, we estimated its area and energy cost by multiplying the values obtained from the memory compiler by 1.5$\times$ to take into account the logic needed to manage bank conflicts and arbitration between read and write ports.

\textbf{Workloads}.
To evaluate the system performance, we adopt two sets of benchmarks.
The first is a synthetic benchmark suite comprising 63 $3{\times}3$ Conv2D layers built using common values for batch size (B), height (H), and width (W) of the \glspl{oFM}, and the number of input and output channels ($C_{in}$, $C_{out}$).
The second is a benchmark suite comprising \clarity{the Conv2D layers of} $7$ state-of-the-art CNN networks to quantify the speed-up and the energy savings on models with different architectures. 
Within the selected benchmarks, ResNet-34 and ResNet-50~\cite{He2015} are taken as representative of computationally intensive networks used for classification tasks; RetinaNet-ResNet50-fpn~\cite{Lin_2017_ICCV}, SSD-VGG16~\cite{ssd300_vgg16}, and YOLOv3~\cite{redmon2018yolov3} for object detection tasks; UNet~\cite{unet} for high-resolution semantic segmentation tasks.
We used the implementation of the networks available in the Torchvision Python package.
\begin{table*}[ht]
\caption{Throughput of the Winograd operator normalized to the im2col operator for different $3{\times}3$ Conv2D layers with stride equals to 1 and padding \textit{same}. $H,W$ refers to the output resolution.} 
 \label{tab:throughput_wino4_norm_im2col}
 \centering
 \begin{minipage}{0.8\textwidth}
    \centering
    \includegraphics[width=\linewidth]{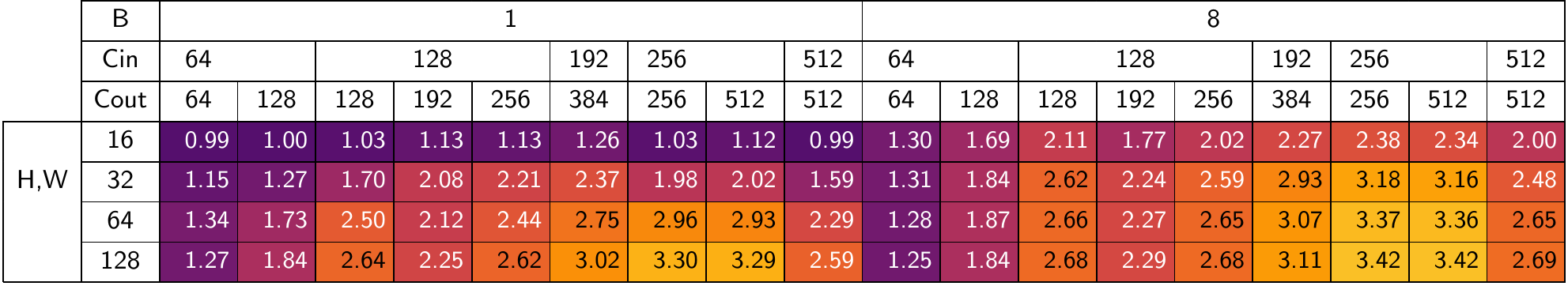}
 \hspace{-1.5cm}
 \end{minipage}%
 \begin{minipage}{0.145\textwidth}
    \centering
    \includegraphics{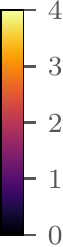}
 \end{minipage}
\end{table*}
 
\input{tables/accelerator_breakdown.tex}

\subsubsection{Area and Power Analysis}
\Cref{tab:area_power_breakdown} reports the detailed area and power breakdown of the AI core and the physical layout of the implemented hardware extensions. The \unit{Cube Unit} dominates both the area and the power of the compute modules, being at least \revised{$6.4\times$} larger and requiring \revised{$6.7\times$} more power than a single Winograd transformation engine.

Overall, all Winograd transformation engines occupy \revised{merely $6.1\%$} of the core area. 
Note that most of the time, only the input and the output transformation engines are active simultaneously, whereas the power cost of the weight transformation engine is amortized over the computations of all activations.

Thus, the Winograd extension adds \revised{$\approx 17\%$} of power overhead to the \unit{Cube Unit}, but it also reduces its number of active cycles to one-fourth compared to the im2col.
\revised{As shown in \Cref{tab:area_power_breakdown}, the power consumption of the \unit{Cube Unit} and of the \unit{L0C-Port B} increases for the Winograd kernel by $1.26\times$ and $2.22\times$, respectively. 
The lower sparsity of activations and weights in the Winograd domain~\cite{liu2018efficientsparse} increases the switching power consumption. 
Nevertheless, the compute datapath is $\approx3\times$ more energy efficient when using the Winograd kernel instead of the im2col}.
On the memory side, both area and energy per access are highly correlated with the size of the memory.
Although \unit{L0A} exposes a more complex access pattern compared to \unit{L0B} (see \cref{sec:xform_engine}), the overhead on area and energy access cost is negligible.
On the other hand, the rotation logic needed on the output port (\textit{PortB}) of \unit{L0C} remarkably affects the power consumption.
However, the (\textit{PortB}) of \unit{L0C} is, on average, less utilized compared to (\textit{PortA}), making the average access cost to \unit{L0C} much less expensive in practice.

\subsubsection{Throughput Analysis}\label{sec:throughputanalysis}

\Cref{tab:throughput_wino4_norm_im2col} shows the speed-up of the Winograd operator compared to the im2col for different parameters of a $3{\times}3$ Conv2D layer.
Although the performance of the Winograd algorithm is highly dependent on the characteristics of the workload, we can identify two macro-trends.

\begin{figure}[t]
    \centering
    \includegraphics[width=0.95\columnwidth]{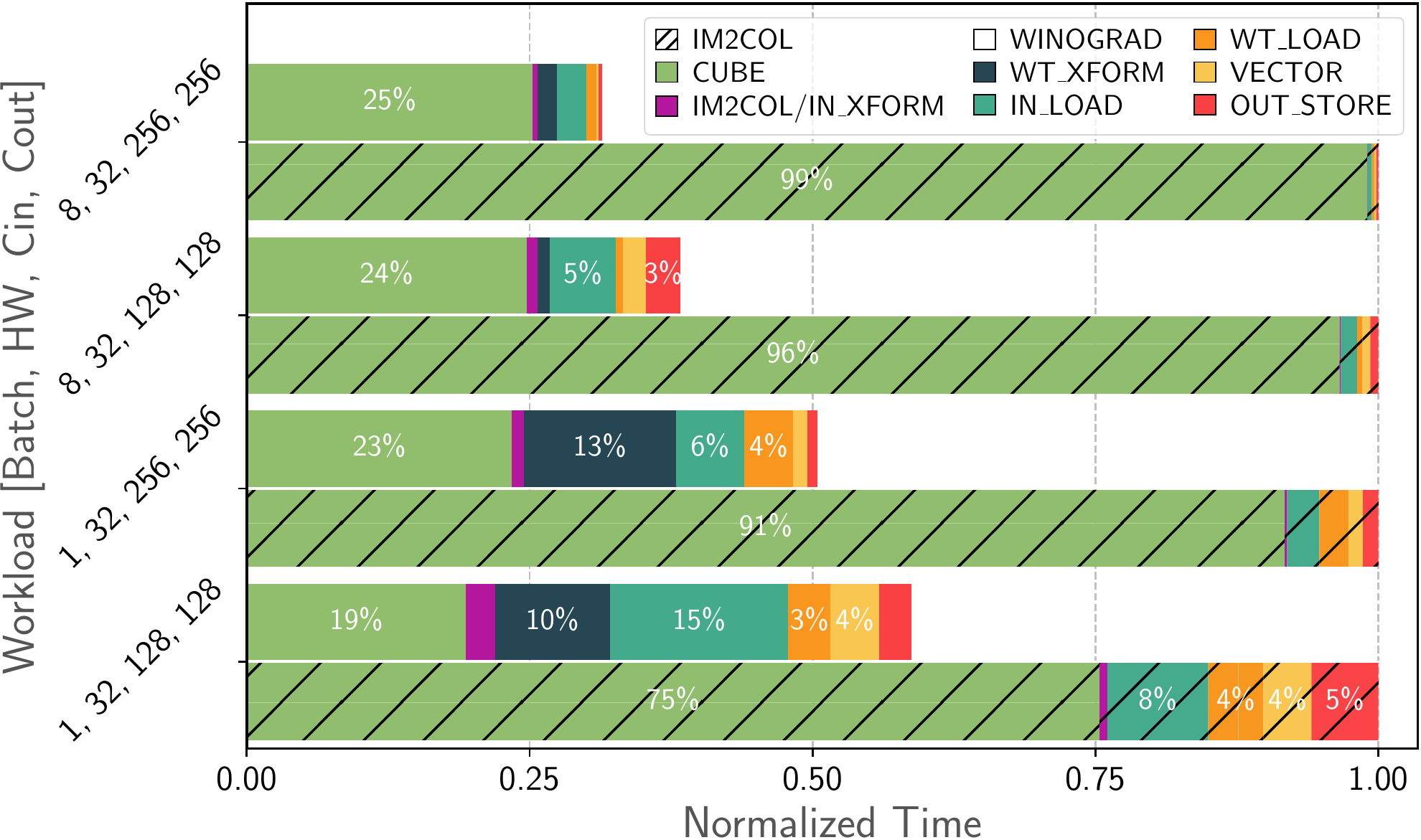}
    \caption{Cycle Breakdown for im2col vs. Winograd $F_4$.}
    \label{fig:critical_path_sampled}
\end{figure}

\textit{Larger resolution or batch size $\rightarrow$ higher speed-up}.
As explained in \cref{sec:winograd_operator}, we adopt a weight-stationary dataflow, where the weights are reused for all the \glspl{iFM}. 
Thus, when the reuse of the weights is small, the performance is limited by the transfer of the weights.
For example, the speed-up increases from \revised{$1.98\times$ to $3.30\times$} when the resolution increases from $32{\times}32$ to $128{\times}128$ with $256$ input and output channels at batch size equal to $1$, and from $1.98\times$ to $3.18\times$ when changing the batch size from $1$ to $8$ at iso-resolution ($32{\times}32$).
To better visualize the bottlenecks for different workloads, \cref{fig:critical_path_sampled} shows the cycle usage breakdown of the critical path of the Winograd operator normalized to the im2col (hatched bar).
Specifically, comparing the first and the third workloads in \cref{fig:critical_path_sampled}, \clarity{a batch size of 8 instead of 1} decreases the normalized percentage of cycles occupied by weight transfer and weight transformations from $13\%$ to $2\%$.
Note that the throughput of the weight transformation engine has been tuned to match the external weight transfers while occupying the minimum area.
Thus, removing the contribution of the weight transformation engine will reveal another critical path where the weight data transfer takes the place of the transformations. 
\revised{This analysis also shows the need for transforming the weights on-the-fly instead of reading the transformed weights from the external memory. As the weights in the Winograd domain are $4\times$ larger than the in the spatial domain, the load overhead will be much higher and difficult to amortize.}

\textit{Larger number of input channels $\rightarrow$ higher speed-up}. 
A larger number of input channels increases the output reuse within the core, reducing the memory bandwidth occupied by the write operations of the \glspl{oFM}.
This increased data reuse frees bandwidth for the transfer of the \glspl{iFM}, with a remarkable effect on performance as the \glspl{iFM} are broadcasted to the two cores.
For example, the speed-up increases from \revised{$2.62\times$} to \revised{$3.18\times$} when increasing the number of input channels from $128$ to $256$, with batch size equal to $8$, spatial resolution equal to $32{\times}32$, and output channels equal to $256$.
In \cref{fig:critical_path_sampled}, having more bandwidth to reserve for the \glspl{iFM} transfer reduces the cycles occupied by the MTE2 from \revised{$5\%$} to \revised{$2\%$} for the first two workloads and from \revised{$15\%$} to \revised{$6\%$} for the last ones.
The lack of bandwidth represents, in fact, the main reason why the $F_4$ Winograd operator does not achieve the theoretical $4\times$ speed-up on our system.
As shown in \cref{fig:critical_path_sampled}, the input and the output transformation engines never become the bottleneck of the operator as their throughput was sized to exactly match the input and output data rate of the Cube Unit. 

\subsubsection{Comparison with NVDLA}
\shepherd{
In \cref{tab:nvdla_comparison}, we compare our accelerator system with the open-source NVDLA accelerator version 1 which supports direct convolution (in FP16 and INT8) and Winograd $F_2$ (FP16 only) with an on-chip memory of 512\,kB per engine\cite{nvdlaprimer}.
As NVDLA does not provide any tools to convert a model that can be used with its compiler into a format accepted by its verification infrastructure, we have modified the Linux driver used in the virtual platform to write out the sequence of reads and writes from/to the control and status registers of the accelerator, which we then use to simulate the RTL for performance benchmarking. 
The results are compared with the expected values for functional correctness.
}

\shepherd{
The results are summarized in \cref{tab:nvdla_comparison}. 
As a single NVDLA core has a peak throughput of 1\,TOp/s at 1\,GHz, we use 8 NVDLA engines to match the peak throughput of our system, namely, 8\,TOp/s. 
We consider two different configurations for the NVDLA-based system: the leftmost column in~\cref{tab:nvdla_comparison} refers to a system with quasi-infinite bandwidth, whereas the middle column refers to a system with 42.7\,Gword/s, i.e., 85.4\,GB/s in FP16, to match the more realistic bandwidth constraints of our system, i.e., 41 Gword/s (41\,GB/s with INT8 for our system, 82\,GB/s with fp16 for NVDLA).
We use words rather than bytes for the iso-bandwidth evaluation, as the public NVLDA version only supports FP16, and it can be expected that the performance scales with the word width. 
Even though NVDLA with quasi-infinite bandwidth gets close to the theoretical 2.25\x{} speed-up, our accelerator still outperforms NVDLA by 21 to 50\%. 
In the more realistic scenario of the system with limited external bandwidth, Winograd convolutional algorithm on NVDLA becomes strongly memory-bound, which significantly reduces its benefits over the direct convolutional algorithm. 
One of the reasons behind this degradation is that NVDLA needs the weights to be transformed offline, increasing the transferred weight volume by $4^2/3^2=1.78$\x. Furthermore, if the input feature maps of a single layer cannot be stored entirely on-chip, they need to be transferred multiple times from external memory, which leads to cases where the Winograd kernel works even worse than the direct convolution. 
Overall, our accelerator system runs between 1.5 and 3.3\x{} faster than NVDLA at the same peak throughput and same external bandwidth, thanks to using Winograd $F_4$ vs. $F_2$, bandwidth optimization through on-the-fly weight transformation, and higher utilization.
}
\subsubsection{Full Network Evaluation}
\label{sec:full_network_eval}

\input{tables/nvdla_comparison}

\Cref{tab:full_network_evaluation} reports the evaluation of the proposed system on various state-of-the-art CNNs.
The $F_4$ Winograd operator increases the throughput of the $3{\times}3$ compute-heavy Conv2D layers by \revised{$1.9\times$} on average and up to \revised{$2.60\times$}.
The gain on the throughput of the entire network depends on the specific architecture.
In fact, the benefits of the Winograd Algorithm are lower for the networks with many $1{\times}1$ convolutions, like ResNet-50, compared to the networks dominated by the $3{\times}3$ convolutions, like U-Net or YOLOv3.
However, the Winograd algorithm becomes remarkably beneficial when increasing the batch size or the input resolution. 
For example, for ResNet-34, the speed-up achieved increases from \revised{$1.07\times$} to \revised{$1.36\times$} when using a batch size of $16$ instead of $1$.
Even more remarkable is the improvement on \revised{SSD-VGG-16 when increasing the batch size, with the speedup going from $1.55\times$ to $1.83\times$}.

\input{tables/full_network_assessment_v4.tex}

In \cref{tab:full_network_evaluation}, we also report the throughput of the $F_2$ Winograd operator implemented following the same methodology described in \cref{sec:hardware_acc} for $F_4$ and the same dataflow reported in \cref{lst:wino_operator}.
When the $2.25\times$ computational reduction introduced by the $F_2$ operator makes the workloads of the layers memory-bound, $F_2$ and $F_4$ achieve similar performance\revised{, although the $F_4$ configuration always outperforms the $F_2$.}
However, when increasing the batch size or the input resolution, or for very compute-heavy networks such as SSD-VGG-16, YOLOv3, and UNet, $F_4$ increases the throughput w.r.t. to $F_2$ up to \revised{$1.4\times$}.
\revised{To highlight the benefits of the Winograd $F_4$ algorithm, \cref{tab:full_network_evaluation} also shows the speed-up w.r.t. to the im2col algorithm for a system with a higher bandwidth ($1.5\times$, i.e., the ratio between a DDR5 and a DDR4 memory).
In this case, while Winograd $F_2$ hits a plateau around $\approx1.8\times$ of speed-up, the Winograd $F_4$ exploits the additional external memory bandwidth to double the end-to-end throughput compared to the baseline.}

\shepherd{
Overall, Winograd $F_4$ outperforms both im2col and $F_2$ in most cases, even though the improvements over $F_2$ are not always remarkable and highly dependent on the shapes of the specific layers of the network.
Specifically, the presence of the Winograd transformation engines not only constrains the dataflow within a single AI core but also limits the feasible loop transformations, e.g., reordering and blocking, that can be applied on the outer loops of the convolution operation.
Moreover, the spatial resolution of the output activation tiles must be a multiple of $4$, limiting the choice of tiling factors and, in some cases, requiring zero-padding and adding ineffective computations.
All these additional constraints affect the data reuse in the core and the access patterns to the external memory, making the bandwidth limitations even more visible.
Further proof is given by the layer-wise analysis in~\cref{tab:full_network_evaluation}, which reveals that not the same layers of the network are mapped either on Winograd $F_2$ or on Winograd $F_4$ depending on the available extension.
For example, for YOLOv3 with an input resolution of $256$ and batch size $1$, the Winograd $F_2$ outperforms Winograd $F_4$ because it is used to process the deep layers of the network where the small spatial resolution ($\leq 16{\times}16$) makes the Winograd $F_4$ perform worse than the im2col algorithm.
However, for the YOLOv3 with an input resolution of $256$ and batch size $8$, Winograd $F_4$ results in a $1.4\times$ higher throughput than Winograd $F_2$ (i.e., with DDR4).
Apart from the throughput gain, which could be lower than the theoretical $4\times$ FLOPs reduction, the Winograd $F_4$ still reduces the utilization of the \gls{GEMM} engine, usually the most power-hungry computational resources. Therefore the overall energy efficiency is improved, which is analyzed in the following paragraphs.
Thus, depending on the application use cases and the area budget, accelerator designers can use the methodology presented in~\cref{sec:hardware_acceleration} to develop the transformation engines for Winograd $F_2$ and to integrate them together with the Winograd $F_4$ ones, allowing the compiler to select the best computational kernel for each layer of the network. 
}

\begin{figure}[t]
    \centering
    \includegraphics[width=0.96\columnwidth]{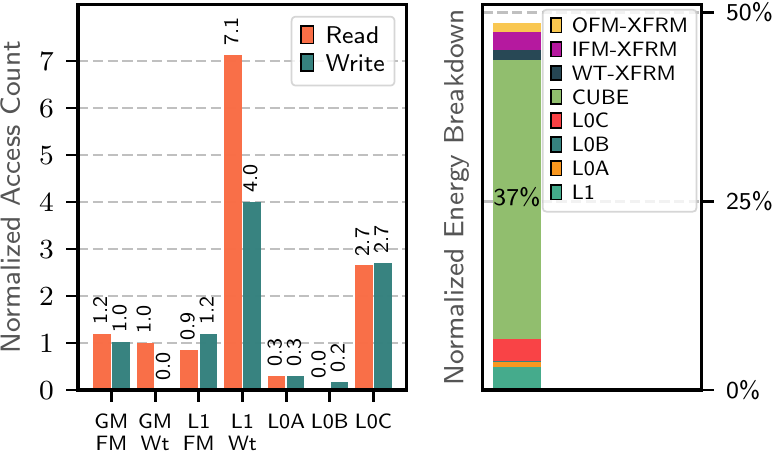}
    \caption{Number of memory accesses (left) and energy breakdown (right) for Winograd $F_4$ w.r.t. im2col.}
    \label{fig:memory_accesses}
\end{figure}

\Cref{fig:memory_accesses} reports, on the left, the average number of read and write accesses and, on the right, the average energy breakdown of the Winograd $F_4$ operator for the Winograd layers only of the networks reported in \cref{tab:full_network_evaluation}.
All values are normalized to the im2col Conv2D operator.
The read accesses to the weights in \unit{GM} are the same as the im2col, as the weights are transformed on the fly in the core.
On the other hand, the write accesses to \unit{L1} increases due to the expansion factor caused by the Winograd transformation, namely, $\frac{(m+2)^2}{9}{=}4$ in the case of $F_4$. 
The read accesses to the weights in \unit{L1} increase significantly, as the \unit{Cube Unit} directly reads the weights from \unit{L1} instead of storing and reusing them in \unit{L0B}, as the im2col operator does. 
Nevertheless, the $F_4$ Winograd algorithm reduces the total number of weight reads to one-fourth, and, as the \unit{L1} energy access cost is only $3\times$ higher compared to that of \unit{L0B}, it also lowers the overall energy consumption. 
All the accesses to \unit{L0B} are due to the weight transformations only, and so its cost is highly amortized over time. 
The read accesses to the \glspl{iFM} in \unit{GM} slightly increase, as the write accesses to \unit{L1}, because  the data reuse factor, i.e., the number of output channels, is limited to 64.
The read accesses to the \glspl{iFM} in \unit{L1} and the write accesses to \unit{L0A} decrease as the Winograd transformation increase the volume of the \glspl{iFM} only by a factor of $\frac{(m+2)^2}{m^2}{=}2.25$ for $m=4$ instead of 9 as the im2col for a $3{\times}3$ convolution. 
As the total number of \unit{Cube Unit} active cycles decreases, so does the number of read accesses to \unit{L0A}.
The number of read and write accesses to \unit{L0C} is higher as the \glspl{oFM} are in the Winograd domain.
Overall, the energy spent on the memory subsystem is comparable between $F_4$ Winograd and the im2col operator, yet the Winograd $F_4$ algorithm lowers more than $2\times$ the total energy consumption as it reduces the active cycles of the \unit{Cube Unit}, which, as shown in \cref{fig:memory_accesses}, dominates the energy consumption of the core. 
\shepherd{This analysis reveals another key advantage of the Winograd $F_4$ algorithm compared to the im2col and the Winograd $F_2$ algorithm}: although the theoretical $4\times$ \glspl{MAC} reduction may not always translate into an equivalent throughput increase, it guarantees a higher energy efficiency, which makes it a perfect fit for inference DSA.

%% file: tables/winograd_quant_resnet34.tex
\begin{table}[t]

\caption{Ablation study for ResNet-34 on ImageNet}


\ifdefined\tableninja \small \setlength{\tabcolsep}{4.5pt}\else \setlength{\tabcolsep}{4pt} Switch \tabularxtabularx \fi
\centering
\footnotesize
\begin{tabularx}{\linewidth}{rcccccccrr}


\toprule
Alg. &  WA &$\odot$ & $2^x$ & $\nabla_{\log_2{t}}$  & {KD} & \verb|int|$n$ & Top-5 & Top-1 & \multicolumn{1}{c}{$\Delta$} \\ \midrule

im2col   &               &            &            &            &            & FP32 & \cellcolor[HTML]{63BE7B}90.7 & \cellcolor[HTML]{63BE7B}72.6 & \cellcolor[HTML]{63BE7B}0.0   \\
im2col   &               &            &            &            &            & 8    & \cellcolor[HTML]{63BE7B}90.7 & \cellcolor[HTML]{63BE7B}72.6 & \cellcolor[HTML]{63BE7B}0.0   \\
$F_2$    & \checkmark       &            &            &            &            & 8    & \cellcolor[HTML]{86CB90}90.1 & \cellcolor[HTML]{91CE96}71.4 & \cellcolor[HTML]{91CE96}-1.2  \\
$F_2$   & \checkmark        &            &            &            &            & 8/10 & \cellcolor[HTML]{63BE7B}90.7 & \cellcolor[HTML]{63BE7B}72.6 & \cellcolor[HTML]{63BE7B}0.0   \\
$F_4$  & \checkmark         &            &            &            & \checkmark & 8    & \cellcolor[HTML]{E1A18D}81.6 & \cellcolor[HTML]{E29E8B}59.0 & \cellcolor[HTML]{E29E8B}-13.6 \\
$F_4$  & \checkmark         &            &            &            & \checkmark & 8/10 & \cellcolor[HTML]{BDDDAF}89.2 & \cellcolor[HTML]{C9DAB0}69.1 & \cellcolor[HTML]{C9DAB0}-3.5  \\
\midrule $F_4$ & \checkmark & \checkmark &            &            &            & 8    & \cellcolor[HTML]{86CB90}90.1 & \cellcolor[HTML]{91CE96}71.4 & \cellcolor[HTML]{91CE96}-1.2  \\
$F_4$  & \checkmark         & \checkmark &            &            &            & 8/10 & \cellcolor[HTML]{69C17F}90.6 & \cellcolor[HTML]{7AC689}72.0 & \cellcolor[HTML]{7AC689}-0.6  \\
$F_4$  & \checkmark         & \checkmark &            &            & \checkmark & 8    & \cellcolor[HTML]{63BE7B}90.7 & \cellcolor[HTML]{67C07E}72.5 & \cellcolor[HTML]{67C07E}-0.1  \\
\midrule $F_4$ & \checkmark & \checkmark & \checkmark &            &            & 8    & \cellcolor[HTML]{92CE96}89.9 & \cellcolor[HTML]{A4D5A1}70.9 & \cellcolor[HTML]{A4D5A1}-1.7  \\
$F_4$  & \checkmark         & \checkmark & \checkmark &            &            & 8/10 & \cellcolor[HTML]{6BC180}90.6 & \cellcolor[HTML]{77C586}72.1 & \cellcolor[HTML]{77C586}-0.5  \\
$F_4$  & \checkmark         & \checkmark & \checkmark & \checkmark &            & 8    & \cellcolor[HTML]{A4D5A0}89.6 & \cellcolor[HTML]{A8D6A3}70.8 & \cellcolor[HTML]{A8D6A3}-1.8  \\
$F_4$ & \checkmark          & \checkmark & \checkmark & \checkmark &            & 8/10 & \cellcolor[HTML]{75C486}90.4 & \cellcolor[HTML]{82C98D}71.8 & \cellcolor[HTML]{82C98D}-0.8  \\
$F_4$ & \checkmark          & \checkmark & \checkmark &            & \checkmark & 8    & \cellcolor[HTML]{8CCD93}90.0 & \cellcolor[HTML]{A4D5A1}70.9 & \cellcolor[HTML]{A4D5A1}-1.7  \\
$F_4$  & \checkmark         & \checkmark & \checkmark &            & \checkmark & 8/10 & \cellcolor[HTML]{63BE7B}90.7 & \cellcolor[HTML]{73C484}72.2 & \cellcolor[HTML]{73C484}-0.4  \\
$F_4$  & \checkmark         & \checkmark & \checkmark & \checkmark & \checkmark & 8    & \cellcolor[HTML]{8CCD93}90.0 & \cellcolor[HTML]{9DD29C}71.1 & \cellcolor[HTML]{9DD29C}-1.5  \\
$F_4$   & \checkmark       & \checkmark & \checkmark & \checkmark & \checkmark & 8/10 & \cellcolor[HTML]{63BE7B}90.7 & \cellcolor[HTML]{73C484}72.3 & \cellcolor[HTML]{73C484}-0.3 \\  \bottomrule  

\end{tabularx}\label{tab:overviewapproaches}
\vspace{-2mm}
\flushleft $\odot$: tap-wise quantization, $2^x$: power-of-two quant.,
$\nabla_{\log_2{t}}$ training, KD: knowledge distillation, WA: Train Winograd-Aware.\\
\end{table}

%% file: tables/winograd_soa.tex
\begin{table}[t]

\caption{SoA Winograd-aware quantization methods.} \label{tab:overviewSoA}
\ifdefined\tableninja \small \setlength{\tabcolsep}{6pt}\else \setlength{\tabcolsep}{4pt} Switch \tabularxtabularx \fi
\footnotesize
\begin{tabular}{lrcrrr}
\toprule   CIFAR-10/ResNet-20                              & & \verb|int|$n$  & Top-1             & Ref. & $\Delta$ \\

\midrule   \cite{barabasz2020quantaized} Legendre (static) & $F_4$  & 8    & 85.0              & 92.3 & -7.3 \\
\cite{barabasz2020quantaized}   Legendre (static)          & $F_4$  & 8/9  & 89.4              & 92.3 & -2.9 \\
\cite{barabasz2020quantaized}   Legendre (flex)            & $F_4$  & 8    & 91.8              & 92.3 & -0.5 \\
\cite{barabasz2020quantaized}   Legendre (flex)            & $F_4$  & 8/9  & 92.3              & 92.3 & \textbf{0.0}  \\
\cite{fernandez2020searching}   Winograd-Aware (s)         & $F_4$  & 8    & 84.3\footnotemark & 93.2 & -8.9 \\
\cite{fernandez2020searching}   Winograd-Aware (f)         & $F_4$  & 8    & 92.5              & 93.2 & -0.7 \\
\cite{li2021winograd} Winograd   AdderNet                  & $F_2$  & 8    & 91.6              & 92.3 & -0.7 \\
Tapwise Quant.                                     & $F_4$  & 8    & 93.8              & 94.4 & -0.6 \\
\vspace{2mm}Tapwise Quant.                                     & $F_4$  & 8/9  & 94.4                & 94.4 & \textbf{0.0}    \\
CIFAR-10/VGG-nagadomi                                      &        & \verb|int|$n$  & Top-1             & Ref. & $\Delta$ \\
\midrule \cite{liu2018efficientsparse} Sparse                    & $F_2$  & FP32 & 93.4              & 93.3 & 0.1 \\
\cite{li2020lance} Quant. Winograd & $F_2$	& 8 &	90.3	&90.4	&-0.1\\

Tapwise Quant. (static)                                    & $F_4$  & 8    & 90.8              & 92.0 & -1.2 \\

Tapwise Quant. (static)                                    & $F_4$  & 8/9    & 91.9              & 92.0 & -0.1 \\
\vspace{2mm}Tapwise Quant. (static)                        & $F_4$  & 8/10 & 92.0              & 92.0 & \textbf{0.0}    \\
ImageNet/ResNet-50                                         &        & \verb|int|$n$  & Top-1             & Ref. & $\Delta$ \\
\midrule\cite{meng2019efficient} Complex   Numbers         & $F_4$  & 8    & 73.2              & 73.3 & -0.1 \\
\cite{liu2020efficientRNS} Residue Numbers                    & $F_{14}$ & 8    & 75.1              & 76.1 & -1.0 \\
\cite{li2021lowino} LoWino	& $F_4$	& FP32/8&	75.5&	76.1&	-0.6 \\

Tapwise Quant. (static)                                    & $F_4$  & 8    & 75.2              & 75.5 & -0.3 \\
Tapwise Quant. (static)                                    & $F_4$  & 8/10 & 75.5              & 75.5 & \textbf{0.0} 

\\ \bottomrule
\end{tabular}
\vspace{-1mm}
\flushleft \textsuperscript{1}Not reported. Reproduced with the open-source code \cite{fernandez2020searching}.
\end{table}

%% file: tables/accelerator_breakdown.tex
\begin{table}[!htpb]
    \footnotesize
    \caption{AI Core breakdown at 0.8\,V and \revised{500}\,MHz. \revised{Power consumptions marked with $^*$ refer to the Im2col kernel, or with $^\dagger$ to the $F_4$ Winograd kernel. The cube TOp/s/W reported for the $F_4$ Winograd kernel are computed using the equivalent TOp in the spatial domain, i.e., $4\times$ the TOp of the \unit{Cube Unit}.}}
    \label{tab:area_power_breakdown}
    \vspace{-0.5cm}
    \begin{center}
    \small
    \setlength{\tabcolsep}{2.8pt}
    \footnotesize
    \begin{tabular}{llr@{\hspace{1mm}}rrr}
    \toprule
    \multicolumn{2}{c}{\textbf{Unit}} & \multicolumn{2}{c}{\textbf{Area}} & \textbf{Peak Power} & \revised{\textbf{TOp/s/W}}\\
    \midrule
     \multirow{2}{*}{Cube} &  & \multirow{2}{*}{$\revised{2.04}\,\si{mm^2}$} & \multirow{2}{*}{$(\revised{19.2}\%)$} & \revised{$^*$1521}\,mW & \revised{$^*$5.39}\\
    &  & & & \revised{$^\dagger$1923}\,mW & \revised{$^+$17.04} \\ \midrule

    \multirow{3}{*}{MTE1} 
    & \small{Im2col} & $\revised{0.03}\,\si{mm^2}$ & $(\revised{0.3}\%)$ & $\revised{30}$\,mW & \\ 
    & IN\_XFORM & $\revised{0.23}\,\si{mm^2}$ & $(\revised{2.2}\%)$ & $\revised{145}$\,mW & \revised{5.3}\\ 
    & WT\_XFORM & $\revised{0.32}\,\si{mm^2}$ & $(\revised{3.0}\%)$ & $\revised{228}$\,mW & \revised{1.6}\\ 
    \midrule 
    FIX\_PIPE & OUT\_XFORM & $\revised{0.10}\,\si{mm^2}$ & $(\revised{0.9}\%)$ & \revised{$114$}\,mW & \revised{2.3}\\ 
    \end{tabular}
    \setlength{\tabcolsep}{5.9pt}
    \begin{tabular}{l@{\hspace{3mm}}r@{\hspace{3mm}}r@{\hspace{1mm}}r@{\hspace{3mm}}r@{\hspace{3mm}}r}
    \toprule
    \textbf{Memory} & \textbf{Size} & \multicolumn{2}{c}{\textbf{Area}} & \textbf{Rd Cost} & \textbf{Wr Cost}\\
    \midrule
    L0A & 64\,kB &  $0.32\,\si{mm^2}$ & $(\revised{3.1}\%)$ & $0.22$\,\si{pJ/B} & $0.24$\,\si{pJ/B} \\ 
    L0B & 64\,kB & $0.32\,\si{mm^2}$ & $(\revised{3.1}\%)$ & $0.22$\,\si{pJ/B} & $0.24$\,\si{pJ/B} \\ 
    L0C - PortA & \multirow{3}{*}{$288$\,kB} & \multirow{3}{*}{$\revised{1.24}\,\si{mm^2}$} & \multirow{3}{*}{$(\revised{11.7}\%)$} & $0.23$\,\si{pJ/B} & $0.29$\,\si{pJ/B} \\
    \multirow{2}{*}{L0C - PortB} & & & & \revised{$^*0.31$}\,\si{pJ/B} & - \\
      & & & & \revised{$^+0.69$}\,\si{pJ/B} & - \\
    L1 
    & \revised{1248}\,kB & \revised{$5.97$}\,\si{mm^2} & $(\revised{56.1}\%)$ & \revised{$0.92$}\,\si{pJ/B} & \revised{$0.68$}\,\si{pJ/B} \\
    \bottomrule
    \end{tabular}
    \includegraphics[width=0.7\columnwidth]{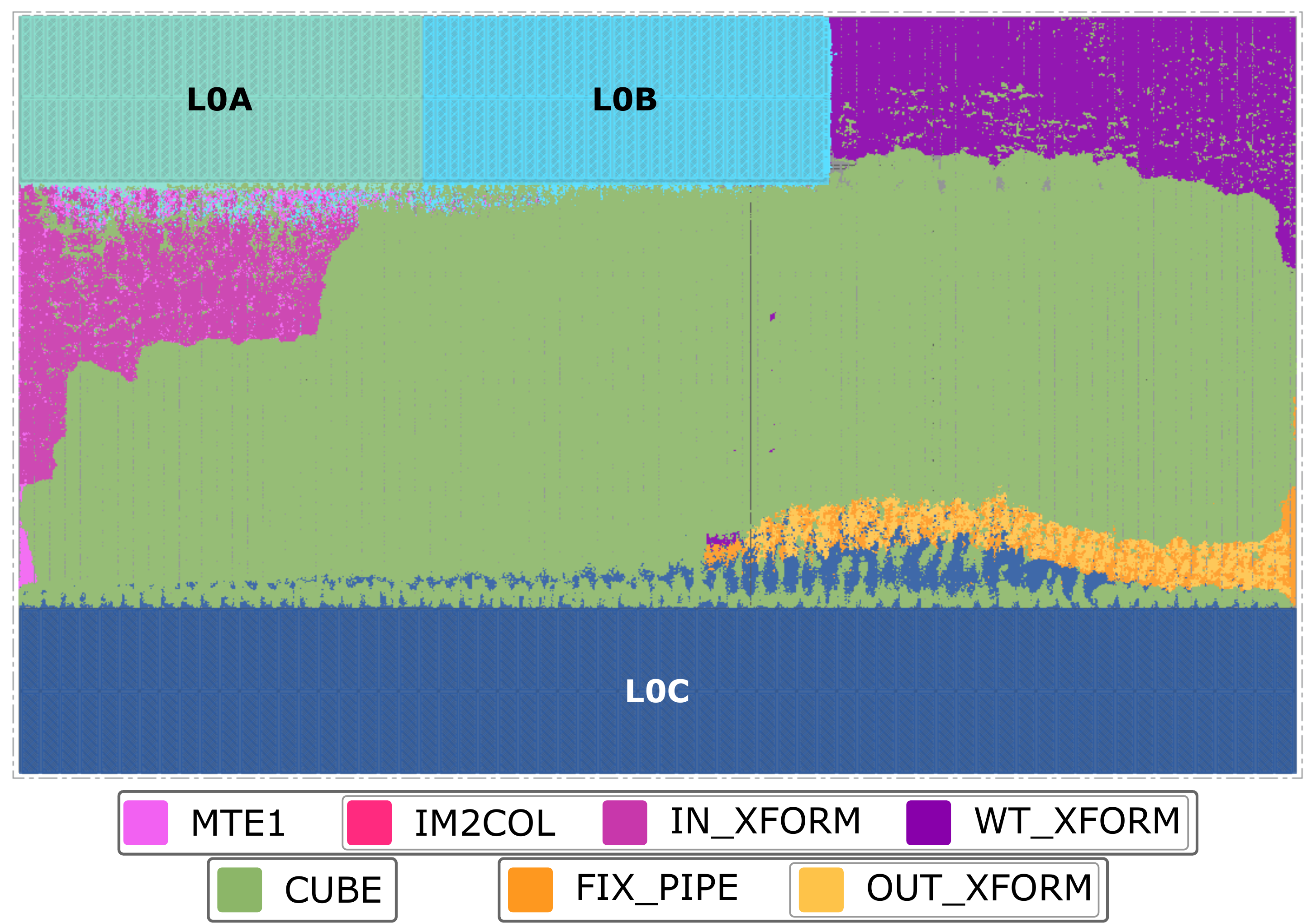}
    \vspace{-3mm}
    \end{center}
 \end{table}

%% file: tables/nvdla_comparison.tex
\begin{table}[t]
\renewcommand{\arraystretch}{1.25}

\caption{\shepherd{Comparison of NVDLA and our accelerator system.}\label{tab:nvdla_comparison}}

\setlength{\tabcolsep}{2.25pt}
\begin{tabular}{r|rrrrrr}
\toprule
 & \multicolumn{2}{c}{\textbf{8\x F2 NVDLA}}           & \multicolumn{2}{c}{\textbf{8\x F2 NVDLA}}           & \multicolumn{2}{c}{\textbf{F4 OURS}}          \\
\multicolumn{1}{c|}{}                                           

Bandwidth\textsuperscript{1} & \multicolumn{2}{c}{128 Gword/s}        & \multicolumn{2}{c}{42.7 Gword/s}      & \multicolumn{2}{c}{41 Gword/s}       \\
Peak Throughput                                                                                                  & \multicolumn{2}{c}{8 TOp/s} & \multicolumn{2}{c}{8 TOp/s} & \multicolumn{2}{c}{8 TOp/s} \\
\textbf{B, H, W, $C_\text{in}$, $C_\text{out}$}
                                                                & t [\textmu s]         & SU [$\times$]       & t [\textmu s]         & SU [$\times$]      & t [\textmu s]        & SU [$\times$]      \\ 
 \hline
\textbf{8, 32, 32, 128, 128}                                                                                               & 79.1                  & 2.03           & 106.2                 & 1.74           & 59.8                & \textbf{2.62}           \\
\textbf{8, 32, 32, 128, 256}                                                                                               & 144.7                 & 2.13           & 175.8                 & 1.89           & 118.7               & \textbf{2.59}           \\
\textbf{8, 32, 32, 256, 512}                                                                                                & 574.6                 & 2.09           & 1736.5                & 0.72           & 383.7                      
            & \textbf{3.16}  \\
\bottomrule
\end{tabular}
\vspace{-1mm}
\flushleft \shepherd{\textsuperscript{1}Word-bandwidth to external memory: 1 word is 2\,Bytes for FP16 (NVDLA) and 1\,Byte for INT8 (ours). We compare NVDLA with quasi-infinite bandwidth (i.e., 256\,GB/s) and iso-word-bandwidth.}
\end{table}

%% file: tables/full_network_assessment_v4.tex
\begin{table*}[!ht]
\begin{center}

     \ifdefined\tableninja \small \setlength{\tabcolsep}{1.75pt}\else \setlength{\tabcolsep}{6pt} \fi
    \footnotesize
    \caption{Throughput and energy efficiency evaluation. Values in parentheses refer only to the Winograd layers. \revised{The speed-up columns marked with the symbol $^*$ refer to a system with a higher external memory bandwidth ($1.5\times$)}.\label{tab:full_network_evaluation}}
    \begin{tabular}{lll|rrrrrrrrr|c}
    \toprule
    &
    &
    &
    \multicolumn{9}{c|}{\textbf{Throughput [Imgs/s]}} &
    \textbf{\begin{tabular}[c]{@{}c@{}}Energy\\Eff.[Inf/J]\end{tabular}} \\ 
    \textbf{Network}    &
    \textbf{Batch}      &
    \textbf{Res.} &
    \textbf{im2col} &
    \textbf{$F_2$} &
    \textbf{$F_4$} &
    \textbf{$F_2$ vs. im2col} &
    \textbf{$F_4$ vs. im2col} &
	\shepherd{\textbf{$F_4$ vs. $F_2$}} &
    \textbf{$^*F_2$ vs. im2col} &
    \textbf{$^*F_4$ vs. im2col} &
	\shepherd{\textbf{$^*F_4$ vs. $F_2$}} &
    \textbf{$F_4$ vs. im2col} \\
        \midrule

              ResNet-34 &      1 &  224 &  921 &  950 & 985  & 1.03x (1.29x) &  1.07x (1.39x) & \shepherd{1.04x (1.08x)} &  1.07x (1.15x) & 1.10x (1.52x) & \shepherd{1.03x (1.32x)} & 1.15x (1.79x) \\
              ResNet-50 &      1 &  224 &  669 &  676 & 684  & 1.01x (1.29x) &  1.02x (1.37x) & \shepherd{1.01x (1.06x)} &  1.02x (1.14x) & 1.03x (1.51x) & \shepherd{1.01x (1.32x)} & 1.05x (1.79x) \\
         RetinaNet-R-50 &      1 &  800 &   38 &   51 &  57  & 1.34x (1.73x) &  1.49x (2.18x) & \shepherd{1.11x (1.26x)} &  1.40x (1.79x) & 1.60x (2.43x) & \shepherd{1.14x (1.36x)} & 1.51x (2.00x) \\
             SSD-VGG-16 &      1 &  300 &  162 &  243 & 252  & 1.50x (1.59x) &  1.55x (1.89x) & \shepherd{1.03x (1.19x)} &  1.58x (1.75x) & 1.71x (2.05x) & \shepherd{1.08x (1.17x)} & 1.70x (2.03x) \\
                   UNet &      1 &  572 &   46 &   75 &  81  & 1.62x (1.71x) &  1.74x (2.18x) & \shepherd{1.07x (1.27x)} &  1.75x (1.85x) & 2.00x (2.49x) & \shepherd{1.14x (1.35x)} & 1.85x (2.28x) \\
                 YOLOv3 &      1 &  256 &  317 &  349 & 358  & 1.10x (1.34x) &  1.13x (1.46x) & \shepherd{1.03x (1.09x)} &  1.15x (1.47x) & 1.16x (1.41x) & \shepherd{1.01x (0.96x)} & 1.43x (2.23x) \\
                 YOLOv3 &      1 &  416 &  154 &  188 & 195  & 1.22x (1.66x) &  1.27x (1.85x) & \shepherd{1.04x (1.11x)} &  1.27x (1.77x) & 1.35x (1.83x) & \shepherd{1.06x (1.03x)} & 1.35x (1.92x) \\
             SSD-VGG-16 &      8 &  300 &  176 &  304 & 328  & 1.68x (1.71x) &  1.83x (1.97x) & \shepherd{1.09x (1.15x)} &  1.74x (1.77x) & 2.06x (2.26x) & \shepherd{1.18x (1.28x)} & 1.78x (1.90x) \\
                 YOLOv3 &      8 &  256 &  496 &  664 & 680  & 1.33x (1.72x) &  1.37x (2.40x) & \shepherd{1.03x (1.40x)} &  1.42x (1.87x) & 1.51x (2.32x) & \shepherd{1.06x (1.24x)} & 1.50x (2.57x) \\
              ResNet-34 &     16 &  224 & 1472 & 1776 & 2000 & 1.22x (1.73x) &  1.36x (1.93x) & \shepherd{1.11x (1.12x)} &  1.24x (1.52x) & 1.46x (2.29x) & \shepherd{1.18x (1.51x)} & 1.40x (2.03x) \\
              ResNet-50 &     16 &  224 &  816 &  848 &  880 & 1.05x (1.73x) &  1.07x (1.90x) & \shepherd{1.02x (1.10x)} &  1.06x (1.51x) & 1.10x (2.25x) & \shepherd{1.04x (1.49x)} & 1.13x (2.02x) \\
                 YOLOv3 &     16 &  256 &  480 &  672 &  672 & 1.38x (1.92x) &  1.38x (2.60x) & \shepherd{1.00x (1.35x)} &  1.44x (2.01x) & 1.51x (2.46x) & \shepherd{1.05x (1.22x)} & 1.51x (2.59x) \\
    \bottomrule
    \end{tabular}
    \end{center}
\end{table*}

%% file: 5_relatedworks.tex
\section{Related Work}
\label{sec:related_work}

\textbf{Winograd Algorithm.} Several works have been proposed to extend the original Winograd algorithm~\cite{winograd1980arithmetic} to work on general 2D convolution~\cite{Lavin2015a, yepez2020stride, yang2020stride}, and to improve its performance by combining it with the Strassen algorithm~\cite{zhao2018faster} or its numerical accuracy by using higher-order polynomials~\cite{barabasz_2019_winogradbeyondlinear} and better polynomial root points for $m>4$~\cite{barabasz2020error, alam2022winograd}.
Li et al.~\cite{li2021winograd} combined the Winograd algorithm with AdderNet, which uses $\ell_1$ instead of $\ell_2$ norm for feature extraction, therefore replacing all MAC operations with additions. 
However, on CIFAR-10/ResNet-20, the proposed method introduces an accuracy drop from 92.3\% for the FP32 baseline to 91.6\%. 
Sparsity has been extensively used to reduce the computational complexity of CNNs.
Liu et al.~\cite{liu2018efficientsparse} and Li et al.~\cite{li2017enabling} proposed to move the ReLU activation layer after the input transformation and to prune the weights in the Winograd domain. 
However, they use FP32 networks and only report the reduction of the number of MACs.
Combining pruning with tap-wise quantization and assessing its benefit on a hardware accelerator represents an interesting future work direction.

\textbf{Quantized Winograd.}
Gong et al.~\cite{JiongGong2018} and Li et al.~\cite{li2020lance} proposed to quantize $F_2$ in the Winograd domain with a single quantization scalar per transformation.
Meng et al.~\cite{meng2019efficient} extended the algorithm to use complex root points, increasing the number of valid root points for Winograd $F_4$. 
Liu et al.~\cite{liu2020efficientRNS} proposed to combine Winograd and Residue Number System (RNS), selecting 8\,bit moduli 251,241,239 and Winograd $F_{14}$. 
Fernandez et al.~\cite{fernandez2020searching} proposed Winograd-Aware Training for Quantized Neural Networks, where gradients are propagated through the Winograd Domain. 
In the case of $F_4$, they had to re-train the transformation matrices (WA-flex), making the transformation operation dense and introducing FP32 MACs.
Barabasz et al.~\cite{barabasz_2019_winogradbeyondlinear} extended the work of Fernandez et al.~\cite{fernandez2020searching} using Legendre polynomial bases, where 6 additional sparse diagonal matrix multiplications are required.
Li et al.~\cite{li2021lowino} proposed to use FP32 feature maps and weights but to quantize the weights and feature maps in the Winograd domain.
In this way, the elementwise multiplication can be performed using \verb|int8|, whereas the input and output transformations are carried out in FP32. 

\textbf{Custom Winograd Accelerators.} 
Several custom accelerators targeting FPGAs were proposed to accelerate the Winograd algorithm~\cite{liu2021winocnn,  lu2018spwa, yang2021biswsrbs}. They comprise a spatial architecture capable of performing only the Winograd algorithm, whereas we propose a methodology to integrate Winograd support in a programmable AI accelerator based on a high-throughput MatMul unit, which is the most adopted solution for ASIC accelerators.
Wang et al.~\cite{wang2021customized} proposed a RISC-V extension to support Winograd transformations efficiently.
Xygkis et al.~\cite{xygkis2018efficient} proposed a solution to map the $F_2$ Winograd operator on a general-purpose edge device with Vector Units. 
\revised{The closest proposal to our work is WinDConv~\cite{mahale2020windconv}, an accelerator based on  NVDLA~\cite{nvdlaprimer} which supports the $F_2$ Winograd operator with a fused datapath.
Unfortunately, a one-to-one comparison is difficult as they targeted a mobile application scenario, they reported a post-synthesis-only evaluation using a much newer technological node, and they did not consider the effects of external memory on the performance. 
However, their Winograd extension leads to an increase in energy efficiency over their baseline of $1.82\times$ in the best case, i.e., considering $100\%$ of utilization, whereas our proposal achieves a $2.1\times$ increase of the energy efficiency on average for 12 state-of-the-art CNNs.
Moreover, they also quantize to 6 bits in the spatial domain, which leads to a higher accuracy drop~\cite{nayak2019bit} than the proposed tap-wise quantization flow}.

\textbf{Winograd SW optimizations.}
Several efficient SW implementations of the Winograd algorithms were recently proposed for GPGPUs\cite{liu2021optimizing, castro2021opencnn, kim2021performance} and CPUs~\cite{li2021optimizing, maji2019efficient}.
All these papers adopt similar loop-level optimizations, such as loop unrolling, parallelization, or vectorization, to make the most of the targeted platforms, whose characteristics and constraints differ significantly from ours. 

%% file: 6_conclusion.tex
\section{Conclusion}
We presented the tap-wise quantization algorithm to enable efficient quantized Winograd on 4\x 4 tiles $F_4$.  
Using 8-bits integers for the feature maps and weights and 10-bits integers in the Winograd domain, the $F_4$ Winograd network achieves the same accuracy as the FP32 baseline for ResNet-20 and VGG-nagadomi on the CIFAR-10 benchmark and for ResNet-50 on the ImageNet classification task. 
The proposed method outperforms the state-of-the-art integer-only and $F_4$-aware quantization methods on all the tested networks and tasks. 
Furthermore, we presented a custom HW extension to efficiently process $F_4$ integer Winograd layers and its integration into an industrial-grade AI accelerator. \shepherd{Our proposed system outperforms NVDLA with its Winograd $F_2$ extension by 1.5 to 3.3\x{} at the same compute throughput and bandwidth constraints due to the higher computational reduction from $F_4$, optimized bandwidth requirements by on-the-fly transformations, and higher utilization thanks to the optimized dataflow.}
The proposed hardware extensions have a small area (\revised{6.1\%} of the core area) and power (\revised{17\%} compared to the MatMul engine) overhead over the baseline architecture while achieving up to \revised{3.42\x{}} speed-up on compute-intensive convolutional layers. 
An extensive evaluation over several state-of-the-art computer-vision benchmarks revealed up to \revised{1.83\x{}} end-to-end inference speed-up and \revised{1.85\x{}} energy efficiency improvement.

%% file: main.bbl
\begin{thebibliography}{10}
\providecommand{\url}[1]{#1}
\csname url@samestyle\endcsname
\providecommand{\newblock}{\relax}
\providecommand{\bibinfo}[2]{#2}
\providecommand{\BIBentrySTDinterwordspacing}{\spaceskip=0pt\relax}
\providecommand{\BIBentryALTinterwordstretchfactor}{4}
\providecommand{\BIBentryALTinterwordspacing}{\spaceskip=\fontdimen2\font plus
\BIBentryALTinterwordstretchfactor\fontdimen3\font minus
  \fontdimen4\font\relax}
\providecommand{\BIBforeignlanguage}[2]{{%
\expandafter\ifx\csname l@#1\endcsname\relax
\typeout{** WARNING: IEEEtranS.bst: No hyphenation pattern has been}%
\typeout{** loaded for the language `#1'. Using the pattern for}%
\typeout{** the default language instead.}%
\else
\language=\csname l@#1\endcsname
\fi
#2}}
\providecommand{\BIBdecl}{\relax}
\BIBdecl

\bibitem{alam2022winograd}
S.~A. Alam, A.~Anderson, B.~Barabasz, and D.~Gregg, ``{Winograd Convolution for
  Deep Neural Networks: Efficient Point Selection},'' \emph{arXiv preprint
  arXiv:2201.10369}, 2022.

\bibitem{barabasz2020quantaized}
B.~Barabasz, ``{Quantized Winograd/Toom-Cook Convolution for DNNs: Beyond
  Canonical Polynomials Base},'' \emph{arXiv preprint arXiv:2004.11077}, 2020.

\bibitem{barabasz2020error}
B.~Barabasz, A.~Anderson, K.~M. Soodhalter, and D.~Gregg, ``{Error analysis and
  improving the accuracy of Winograd convolution for deep neural networks},''
  \emph{ACM Transactions on Mathematical Software (TOMS)}, vol.~46, no.~4, pp.
  1--33, 2020.

\bibitem{barabasz_2019_winogradbeyondlinear}
B.~Barabasz and D.~Gregg, ``{Winograd Convolution for DNNs: Beyond Linear
  Polynomials},'' in \emph{AI*IA 2019 -- Advances in Artificial Intelligence},
  M.~Alviano, G.~Greco, and F.~Scarcello, Eds.\hskip 1em plus 0.5em minus
  0.4em\relax Cham: Springer International Publishing, 2019, pp. 307--320.

\bibitem{bengio2013estimating}
Y.~Bengio, N.~L{\'e}onard, and A.~Courville, ``Estimating or propagating
  gradients through stochastic neurons for conditional computation,''
  \emph{arXiv:1308.3432}, 2013.

\bibitem{blahut2010fast}
R.~E. Blahut, \emph{{Fast algorithms for signal processing}}.\hskip 1em plus
  0.5em minus 0.4em\relax Cambridge University Press, 2010.

\bibitem{castro2021opencnn}
R.~L. Castro, D.~Andrade, and B.~B. Fraguela, ``{OpenCNN: A Winograd Minimal
  Filtering Algorithm Implementation in CUDA},'' \emph{Mathematics}, vol.~9,
  no.~17, p. 2033, 2021.

\bibitem{im2col_paper}
\BIBentryALTinterwordspacing
K.~Chellapilla, S.~Puri, and P.~Simard, ``{High Performance Convolutional
  Neural Networks for Document Processing},'' in \emph{{Tenth International
  Workshop on Frontiers in Handwriting Recognition}}, G.~Lorette, Ed.,
  {Universit{\'e} de Rennes 1}.\hskip 1em plus 0.5em minus 0.4em\relax La Baule
  (France): {Suvisoft}, Oct. 2006, http://www.suvisoft.com. [Online].
  Available: \url{https://hal.inria.fr/inria-00112631}
\BIBentrySTDinterwordspacing

\bibitem{Chen2016}
Y.-H. Chen, T.~Krishna, J.~Emer, and V.~Sze, ``{Eyeriss: An Energy-Efficient
  Reconfigurable Accelerator for Deep Convolutional Neural Networks},'' in
  \emph{Proc. IEEE ISSCC}, 2016, pp. 262--263.

\bibitem{nvidiaa100}
J.~Choquette, W.~Gandhi, O.~Giroux, N.~Stam, and R.~Krashinsky, ``Nvidia a100
  tensor core gpu: Performance and innovation,'' \emph{IEEE Micro}, vol.~41,
  no.~2, pp. 29--35, 2021.

\bibitem{fernandez2020searching}
J.~Fernandez-Marques, P.~N. Whatmough, A.~Mundy, and M.~Mattina, ``{Searching
  for winograd-aware quantized networks},'' \emph{MLSys}, 2021.

\bibitem{guo2018survey}
Y.~Guo, ``{A survey on methods and theories of quantized neural networks},''
  \emph{arXiv preprint arXiv:1808.04752}, 2018.

\bibitem{hackenberg2015energy}
D.~Hackenberg, R.~Sch{\"o}ne, T.~Ilsche, D.~Molka, J.~Schuchart, and R.~Geyer,
  ``An energy efficiency feature survey of the intel haswell processor,'' in
  \emph{2015 IEEE international parallel and distributed processing symposium
  workshop}.\hskip 1em plus 0.5em minus 0.4em\relax IEEE, 2015, pp. 896--904.

\bibitem{Han2015}
S.~Han, J.~Pool, m.~j. Tran, W.~J. Dally, J.~Tran, W.~J. Dally, m.~j. Tran, and
  W.~J. Dally, ``{Learning Both Weights and Connections for Efficient Neural
  Networks},'' \emph{NIPS}, 2015.

\bibitem{He2015}
K.~He, X.~Zhang, S.~Ren, and J.~Sun, ``{Deep Residual Learning for Image
  Recognition},'' \emph{Proc. IEEE CVPR}, pp. 770--778, 2015.

\bibitem{hinton2015distilling}
\BIBentryALTinterwordspacing
G.~Hinton, O.~Vinyals, and J.~Dean, ``{Distilling the Knowledge in a Neural
  Network},'' \emph{arXiv preprint arXiv:1503.02531}, 2015. [Online].
  Available: \url{http://arxiv.org/abs/1503.02531}
\BIBentrySTDinterwordspacing

\bibitem{horowitz2014energy}
M.~Horowitz, ``{1.1 computing's energy problem (and what we can do about
  it)},'' in \emph{2014 IEEE International Solid-State Circuits Conference
  Digest of Technical Papers (ISSCC)}.\hskip 1em plus 0.5em minus 0.4em\relax
  IEEE, 2014, pp. 10--14.

\bibitem{howard2017mobilenets}
A.~G. Howard, M.~Zhu, B.~Chen, D.~Kalenichenko, W.~Wang, T.~Weyand,
  M.~Andreetto, and H.~Adam, ``{Mobilenets: Efficient convolutional neural
  networks for mobile vision applications},'' \emph{arXiv preprint
  arXiv:1704.04861}, 2017.

\bibitem{jain2019trained}
S.~R. Jain, A.~Gural, M.~Wu, and C.~H. Dick, ``{Trained quantization thresholds
  for accurate and efficient fixed-point inference of deep neural networks},''
  \emph{arXiv preprint arXiv:1903.08066}, 2019.

\bibitem{JiongGong2018}
{Jiong Gong}, H.~SHEN, X.~D. Lin, and X.~Liu, ``{Method and apparatus for
  keeping statistical inference accuracy with 8-bit winograd convolution},''
  2018.

\bibitem{TPU2016}
\BIBentryALTinterwordspacing
N.~Jouppi, ``{Google supercharges machine learning tasks with TPU custom
  chip},'' 2016. [Online]. Available:
  \url{https://cloudplatform.googleblog.com/2016/05/Google-supercharges-machine-learning-tasks-with-custom-chip.html}
\BIBentrySTDinterwordspacing

\bibitem{kang2021benchmarking}
P.~Kang and J.~Jo, ``{Benchmarking modern edge devices for ai applications},''
  \emph{IEICE Transactions on Information and Systems}, vol. 104, no.~3, pp.
  394--403, 2021.

\bibitem{kim2021performance}
S.~Kim, G.~Park, and Y.~Yi, ``{Performance Evaluation of INT8 Quantized
  Inference on Mobile GPUs},'' \emph{IEEE Access}, vol.~9, pp.
  164\,245--164\,255, 2021.

\bibitem{kingma2014adam}
D.~P. Kingma and J.~Ba, ``{Adam: A method for stochastic optimization},''
  \emph{arXiv preprint arXiv:1412.6980}, 2014.

\bibitem{krishnamoorthi2018quantizing}
R.~Krishnamoorthi, ``Quantizing deep convolutional networks for efficient
  inference: A whitepaper,'' \emph{arXiv preprint arXiv:1806.08342}, 2018.

\bibitem{krizhevsky2009learning}
A.~Krizhevsky and G.~Hinton, ``{Learning multiple layers of features from tiny
  images},'' 2009.

\bibitem{Krizhevsky2012a}
A.~Krizhevsky, I.~Sutskever, and G.~E. Hinton, ``{Imagenet Classification With
  Deep Convolutional Neural Networks},'' in \emph{Adv. NIPS}, 2012.

\bibitem{maeri}
\BIBentryALTinterwordspacing
H.~Kwon, A.~Samajdar, and T.~Krishna, ``Maeri: Enabling flexible dataflow
  mapping over dnn accelerators via reconfigurable interconnects,''
  \emph{SIGPLAN Not.}, vol.~53, no.~2, p. 461–475, mar 2018. [Online].
  Available: \url{https://doi.org/10.1145/3296957.3173176}
\BIBentrySTDinterwordspacing

\bibitem{Lavin2015a}
A.~Lavin and S.~Gray, ``{Fast Algorithms for Convolutional Neural Networks},''
  in \emph{Proc. IEEE CVPR}, 2016, pp. 4013--4021.

\bibitem{li2021optimizing}
D.~Li, D.~Huang, Z.~Chen, and Y.~Lu, ``{Optimizing Massively Parallel Winograd
  Convolution on ARM Processor},'' in \emph{50th International Conference on
  Parallel Processing}, 2021, pp. 1--12.

\bibitem{li2021lowino}
G.~Li, Z.~Jia, X.~Feng, and Y.~Wang, ``{LoWino: Towards Efficient Low-Precision
  Winograd Convolutions on Modern CPUs},'' in \emph{50th International
  Conference on Parallel Processing}, 2021, pp. 1--11.

\bibitem{li2020lance}
G.~Li, L.~Liu, X.~Wang, X.~Ma, and X.~Feng, ``{Lance: efficient low-precision
  quantized winograd convolution for neural networks based on graphics
  processing units},'' in \emph{ICASSP 2020-2020 IEEE International Conference
  on Acoustics, Speech and Signal Processing (ICASSP)}.\hskip 1em plus 0.5em
  minus 0.4em\relax IEEE, 2020, pp. 3842--3846.

\bibitem{li2017enabling}
S.~Li, J.~Park, and P.~T.~P. Tang, ``{Enabling sparse winograd convolution by
  native pruning},'' \emph{arXiv preprint arXiv:1702.08597}, 2017.

\bibitem{li2021winograd}
W.~Li, H.~Chen, M.~Huang, X.~Chen, C.~Xu, and Y.~Wang, ``{Winograd Algorithm
  for AdderNet},'' in \emph{International Conference on Machine
  Learning}.\hskip 1em plus 0.5em minus 0.4em\relax PMLR, 2021, pp. 6307--6315.

\bibitem{liang2021pruning}
T.~Liang, J.~Glossner, L.~Wang, S.~Shi, and X.~Zhang, ``Pruning and
  quantization for deep neural network acceleration: A survey,''
  \emph{Neurocomputing}, vol. 461, pp. 370--403, 2021.

\bibitem{liao2019davinci}
H.~Liao, J.~Tu, J.~Xia, and X.~Zhou, ``Davinci: A scalable architecture for
  neural network computing.'' in \emph{Hot Chips Symposium}, 2019, pp. 1--44.

\bibitem{Lin_2017_ICCV}
T.-Y. Lin, P.~Goyal, R.~Girshick, K.~He, and P.~Dollar, ``Focal loss for dense
  object detection,'' in \emph{Proceedings of the IEEE International Conference
  on Computer Vision (ICCV)}, Oct 2017.

\bibitem{liu2021optimizing}
J.~Liu, D.~Yang, and J.~Lai, ``{Optimizing Winograd-Based Convolution with
  Tensor Cores},'' in \emph{50th International Conference on Parallel
  Processing}, 2021, pp. 1--10.

\bibitem{ssd300_vgg16}
W.~Liu, D.~Anguelov, D.~Erhan, C.~Szegedy, S.~Reed, C.-Y. Fu, and A.~C. Berg,
  ``Ssd: Single shot multibox detector,'' in \emph{Computer Vision -- ECCV
  2016}, B.~Leibe, J.~Matas, N.~Sebe, and M.~Welling, Eds.\hskip 1em plus 0.5em
  minus 0.4em\relax Cham: Springer International Publishing, 2016, pp. 21--37.

\bibitem{liu2018efficientsparse}
X.~Liu, J.~Pool, S.~Han, and W.~J. Dally, ``{Efficient sparse-winograd
  convolutional neural networks},'' \emph{arXiv preprint arXiv:1802.06367},
  2018.

\bibitem{liu2021winocnn}
X.~Liu, Y.~Chen, C.~Hao, A.~Dhar, and D.~Chen, ``{WinoCNN: Kernel sharing
  Winograd systolic array for efficient convolutional neural network
  acceleration on FPGAs},'' in \emph{2021 IEEE 32nd International Conference on
  Application-specific Systems, Architectures and Processors (ASAP)}.\hskip 1em
  plus 0.5em minus 0.4em\relax IEEE, 2021, pp. 258--265.

\bibitem{optimizing_cnn_model_cpus_usenix19}
\BIBentryALTinterwordspacing
Y.~Liu, Y.~Wang, R.~Yu, M.~Li, V.~Sharma, and Y.~Wang, ``Optimizing {CNN} model
  inference on {CPUs},'' in \emph{2019 USENIX Annual Technical Conference
  (USENIX ATC 19)}.\hskip 1em plus 0.5em minus 0.4em\relax Renton, WA: USENIX
  Association, Jul. 2019, pp. 1025--1040. [Online]. Available:
  \url{https://www.usenix.org/conference/atc19/presentation/liu-yizhi}
\BIBentrySTDinterwordspacing

\bibitem{liu2020efficientRNS}
Z.-G. Liu and M.~Mattina, ``{Efficient Residue Number System Based Winograd
  Convolution},'' in \emph{European Conference on Computer Vision}.\hskip 1em
  plus 0.5em minus 0.4em\relax Springer, 2020, pp. 53--68.

\bibitem{lu2018spwa}
L.~Lu and Y.~Liang, ``{SpWA: An efficient sparse winograd convolutional neural
  networks accelerator on FPGAs},'' in \emph{Proceedings of the 55th Annual
  Design Automation Conference}, 2018, pp. 1--6.

\bibitem{mahale2020windconv}
G.~Mahale, P.~Udupa, K.~K. Chandrasekharan, and S.~Lee, ``{WinDConv: A Fused
  Datapath CNN Accelerator for Power-Efficient Edge Devices},'' \emph{IEEE
  Transactions on Computer-Aided Design of Integrated Circuits and Systems},
  vol.~39, no.~11, pp. 4278--4289, 2020.

\bibitem{maji2019efficient}
P.~Maji, A.~Mundy, G.~Dasika, J.~Beu, M.~Mattina, and R.~Mullins, ``{Efficient
  winograd or cook-toom convolution kernel implementation on widely used mobile
  cpus},'' in \emph{2019 2nd Workshop on Energy Efficient Machine Learning and
  Cognitive Computing for Embedded Applications (EMC2)}.\hskip 1em plus 0.5em
  minus 0.4em\relax IEEE, 2019, pp. 1--5.

\bibitem{meng2019efficient}
L.~Meng and J.~Brothers, ``{Efficient winograd convolution via integer
  arithmetic},'' \emph{arXiv preprint arXiv:1901.01965}, 2019.

\bibitem{vta_tvm}
T.~Moreau, T.~Chen, L.~Vega, J.~Roesch, E.~Yan, L.~Zheng, J.~Fromm, Z.~Jiang,
  L.~Ceze, C.~Guestrin, and A.~Krishnamurthy, ``A hardware–software blueprint
  for flexible deep learning specialization,'' \emph{IEEE Micro}, vol.~39,
  no.~5, pp. 8--16, 2019.

\bibitem{Nagadomi2014}
\BIBentryALTinterwordspacing
{Nagadomi}, \emph{{kaggle-cifar10-torch7}}, 2014. [Online]. Available:
  \url{https://github.com/nagadomi/kaggle-cifar10-torch7}
\BIBentrySTDinterwordspacing

\bibitem{nayak2019bit}
P.~Nayak, D.~Zhang, and S.~Chai, ``Bit efficient quantization for deep neural
  networks,'' in \emph{2019 Fifth Workshop on Energy Efficient Machine Learning
  and Cognitive Computing-NeurIPS Edition (EMC2-NIPS)}.\hskip 1em plus 0.5em
  minus 0.4em\relax IEEE, 2019, pp. 52--56.

\bibitem{norrie2021tpuv2v3}
T.~Norrie, N.~Patil, D.~H. Yoon, G.~Kurian, S.~Li, J.~Laudon, C.~Young,
  N.~Jouppi, and D.~Patterson, ``The design process for google's training
  chips: Tpuv2 and tpuv3,'' \emph{IEEE Micro}, vol.~41, no.~2, pp. 56--63,
  2021.

\bibitem{nvdlaprimer}
\BIBentryALTinterwordspacing
NVIDIA. Nvdla primer - nvdla documentation. [Online]. Available:
  \url{http://nvdla.org/primer.html}
\BIBentrySTDinterwordspacing

\bibitem{redmon2018yolov3}
\BIBentryALTinterwordspacing
J.~Redmon and A.~Farhadi, ``{YOLOv3: An Incremental Improvement},''
  \emph{arXiv:1804.02767}, 2018. [Online]. Available:
  \url{http://arxiv.org/abs/1804.02767}
\BIBentrySTDinterwordspacing

\bibitem{unet}
O.~Ronneberger, P.~Fischer, and T.~Brox, ``U-net: Convolutional networks for
  biomedical image segmentation,'' in \emph{Medical Image Computing and
  Computer-Assisted Intervention -- MICCAI 2015}, N.~Navab, J.~Hornegger, W.~M.
  Wells, and A.~F. Frangi, Eds.\hskip 1em plus 0.5em minus 0.4em\relax Cham:
  Springer International Publishing, 2015, pp. 234--241.

\bibitem{Russakovsky2014}
O.~Russakovsky, J.~Deng, H.~Su, J.~Krause, S.~Satheesh, S.~Ma, Z.~Huang,
  A.~Karpathy, A.~Khosla, M.~Bernstein, A.~C. Berg, and L.~Fei-Fei, ``{ImageNet
  Large Scale Visual Recognition Challenge},'' \emph{IJCV}, vol. 115, no.~3,
  pp. 211--252, 2015.

\bibitem{saglietti2022solvable}
L.~Saglietti and L.~Zdeborov{\'a}, ``Solvable model for inheriting the
  regularization through knowledge distillation,'' in \emph{Mathematical and
  Scientific Machine Learning}.\hskip 1em plus 0.5em minus 0.4em\relax PMLR,
  2022, pp. 809--846.

\bibitem{dae_arch}
J.~E. Smith, ``Decoupled access/execute computer architectures,'' in
  \emph{Proceedings of the 9th Annual Symposium on Computer Architecture}, ser.
  ISCA '82.\hskip 1em plus 0.5em minus 0.4em\relax Washington, DC, USA: IEEE
  Computer Society Press, 1982, p. 112–119.

\bibitem{steiner2021exploration}
L.~Steiner, M.~Jung, and N.~Wehn, ``Exploration of ddr5 with the open-source
  simulator dramsys,'' in \emph{MBMV 2021; 24th Workshop}.\hskip 1em plus 0.5em
  minus 0.4em\relax VDE, 2021, pp. 1--11.

\bibitem{villa2021need}
O.~Villa, D.~Lustig, Z.~Yan, E.~Bolotin, Y.~Fu, N.~Chatterjee, N.~Jiang, and
  D.~Nellans, ``Need for speed: Experiences building a trustworthy system-level
  gpu simulator,'' in \emph{2021 IEEE International Symposium on
  High-Performance Computer Architecture (HPCA)}, 2021, pp. 868--880.

\bibitem{wang2021customized}
S.~Wang, J.~Zhu, Q.~Wang, C.~He, and T.~T. Ye, ``{Customized Instruction on
  RISC-V for Winograd-Based Convolution Acceleration},'' in \emph{2021 IEEE
  32nd International Conference on Application-specific Systems, Architectures
  and Processors (ASAP)}.\hskip 1em plus 0.5em minus 0.4em\relax IEEE, 2021,
  pp. 65--68.

\bibitem{streaming_nowatzki}
Z.~Wang and T.~Nowatzki, ``Stream-based memory access specialization for
  general purpose processors,'' in \emph{2019 ACM/IEEE 46th Annual
  International Symposium on Computer Architecture (ISCA)}, 2019, pp. 736--749.

\bibitem{winograd1980arithmetic}
S.~Winograd, \emph{{Arithmetic complexity of computations}}.\hskip 1em plus
  0.5em minus 0.4em\relax Siam, 1980, vol.~33.

\bibitem{xie2017aggregated}
S.~Xie, R.~Girshick, P.~Doll{\'{a}}r, Z.~Tu, and K.~He, ``{Aggregated residual
  transformations for deep neural networks},'' in \emph{Proceedings of the IEEE
  conference on computer vision and pattern recognition}, 2017, pp. 1492--1500.

\bibitem{xygkis2018efficient}
A.~Xygkis, D.~Soudris, L.~Papadopoulos, S.~Yous, and D.~Moloney, ``{Efficient
  winograd-based convolution kernel implementation on edge devices},'' in
  \emph{2018 55th ACM/ESDA/IEEE Design Automation Conference (DAC)}.\hskip 1em
  plus 0.5em minus 0.4em\relax IEEE, 2018, pp. 1--6.

\bibitem{yang2020stride}
C.~Yang, Y.~Wang, X.~Wang, and L.~Geng, ``{A stride-based convolution
  decomposition method to stretch CNN acceleration algorithms for efficient and
  flexible hardware implementation},'' \emph{IEEE Transactions on Circuits and
  Systems I: Regular Papers}, vol.~67, no.~9, pp. 3007--3020, 2020.

\bibitem{yang2021biswsrbs}
T.~Yang, Z.~He, T.~Kou, Q.~Li, Q.~Han, H.~Yu, F.~Liu, Y.~Liang, and L.~Jiang,
  ``{BISWSRBS: A Winograd-based CNN Accelerator with a Fine-grained Regular
  Sparsity Pattern and Mixed Precision Quantization},'' \emph{ACM Transactions
  on Reconfigurable Technology and Systems (TRETS)}, vol.~14, no.~4, pp. 1--28,
  2021.

\bibitem{yepez2020stride}
J.~Yepez and S.-B. Ko, ``{Stride 2 1-D, 2-D, and 3-D winograd for convolutional
  neural networks},'' \emph{IEEE Transactions on Very Large Scale Integration
  (VLSI) Systems}, vol.~28, no.~4, pp. 853--863, 2020.

\bibitem{Zhang2017}
X.~Zhang, X.~Zhou, M.~Lin, and J.~Sun, ``{ShuffleNet: An Extremely Efficient
  Convolutional Neural Network for Mobile Devices},'' in \emph{Proceedings of
  the IEEE Computer Society Conference on Computer Vision and Pattern
  Recognition}, 2018, pp. 6848--6856.

\bibitem{zhao2018faster}
Y.~Zhao, D.~Wang, L.~Wang, and P.~Liu, ``{A faster algorithm for reducing the
  computational complexity of convolutional neural networks},''
  \emph{Algorithms}, vol.~11, no.~10, p. 159, 2018.

\bibitem{zhu2020towards}
F.~Zhu, R.~Gong, F.~Yu, X.~Liu, Y.~Wang, Z.~Li, X.~Yang, and J.~Yan, ``{Towards
  unified int8 training for convolutional neural network},'' in
  \emph{Proceedings of the IEEE/CVF Conference on Computer Vision and Pattern
  Recognition}, 2020, pp. 1969--1979.

\end{thebibliography}
